\documentclass[a4paper, twocolumn,superscriptaddress,preprintnumbers,longbibliography,nofootinbib,allowtoday,accepted=2026-04-30]{quantumarticle}
\pdfoutput=1

\usepackage[utf8]{inputenc}
\usepackage[english]{babel}
\usepackage[T1]{fontenc}
\usepackage[numbers,sort&compress]{natbib}
\usepackage[utf8]{inputenc}
\usepackage{wrapfig}

\usepackage{graphicx}
\usepackage{dcolumn}   % needed for some tables
\usepackage{bm}        % for math
\usepackage{bbm}
\usepackage{dsfont}
\usepackage{mathtools,amssymb,amsmath, amsthm,stmaryrd,thmtools,braket}
\usepackage{empheq}
\usepackage[colorlinks=true, urlcolor=blue,citecolor=blue,anchorcolor=blue]{hyperref}
\usepackage[capitalise]{cleveref}
\usepackage{booktabs}

\newcommand{\tinyspace}{\mspace{1mu}}

\usepackage{tikz}
\usetikzlibrary{quantikz2}
\usepackage{adjustbox}

\newcommand{\norm}[1]{\left\lVert\tinyspace#1\tinyspace\right\rVert}

\usepackage{mathtools}

\newcommand{\Ztwo}{\mathbb{Z}_{2}}

\usepackage{xcolor}

\renewcommand{\arraystretch}{1.5}

\newcommand{\Fig}[1]{Fig.~\ref{#1}}

\usepackage{bm}
\let\vec\bm 

\newcommand{\dual}{\mathrm{dual}}

\let\oldsquare\square
\renewcommand{\square}{{\scriptscriptstyle\oldsquare}}

\begin{document}
\title{Classical shadows for sample-efficient measurements of gauge-invariant observables}
\author{Jacob Bringewatt$^*$}
\affiliation{Volgenau Department of Physics, United States Naval Academy, Annapolis, MD 21402, USA}
\affiliation{Department of Physics, Harvard University, Cambridge, MA 02138, USA}
\author{Henry Froland$^*$}
\affiliation{InQubator for Quantum Simulation (IQuS), Department of Physics,University of Washington, Seattle, WA 98195, USA}
\author{Andreas Elben}
\affiliation{PSI Center for Scientific Computing, Theory and Data, Paul Scherrer Institute, 5232 Villigen PSI, Switzerland}
\affiliation{ETH Zürich - PSI Quantum Computing Hub, Paul Scherrer Institute, 5232 Villigen PSI, Switzerland}
\author{Niklas Mueller}
\affiliation{Center for Quantum Information and Control, University of New Mexico, Albuquerque, NM 87106, USA}
\affiliation{Department of Physics and Astronomy, University of New Mexico, Albuquerque, NM 87106, USA}
\date{\today}
% \maketitle

\begin{abstract}
Classical shadows provide a versatile  framework for estimating many properties of quantum states from repeated, randomly chosen measurements without requiring full quantum state tomography. When prior information is available, such as knowledge of symmetries of states and operators, this knowledge can be exploited to significantly improve sample efficiency. In this work, we develop three classical shadow protocols for $\mathbb{Z}_2$ lattice gauge theory, where a dual formulation enables a rigorous analysis of resource requirements, including both circuit depth and sample complexity. Our approaches can offer exponential improvements in sample complexity over symmetry-agnostic methods, albeit at the cost of increased circuit complexity. While our analysis is restricted to $\mathbb{Z}_2$ lattice gauge theory, our approach offers a blueprint for similar protocols for more general lattice gauge theory models which are currently at the forefront of quantum simulation efforts.
\end{abstract}

\maketitle

\def\thefootnote{*}\footnotetext{These authors contributed equally to this work.}\def\thefootnote{\arabic{footnote}}

\section{Introduction}
Programmable quantum simulators and computers enable the investigation of large many-body systems, for instance, opening new routes for studying systems out-of-equilibrium~\cite{eisert2015quantum,schachenmayer2015thermalization,nandkishore2015many,kaufman2016quantum,daley2022practical,mueller2025quantum,de2024observation,cochran2024visualizing,gonzalez2025observation} or  discover emergent phases of matter~\cite{kandala2017hardware,scholl2021quantum,semeghini2021probing,satzinger2021realizing} that elude classical algorithms.
A key point is that one can not only simulate such complex quantum states, but also characterize them at scale. 
This includes both the extraction of observables to study the properties of these systems~\cite{aaronson2018shadow,huang2020predicting}, but also verification and benchmarking of the computational devices themselves~\cite{emerson2005scalable,knill2008randomized,cramer2010efficient,magesan2011scalable,boixo2018characterizing,erhard2019characterizing,kokail2019self,blume2025quantum}.

A broad range of methods have been developed to characterize quantum states, from tomography protocols that aim to reconstruct the full state~\cite{czerwinski2022selected,cramer2010efficient,gross2010quantum,flammia2012quantum,haah2016sample,o2016efficient,torlai2018neural,saxena2024boundary} to randomized measurement schemes~\cite{van2012measuring,elben2018renyi,vermersch2018unitary,elben2019statistical,brydges2019probing,du2025optimal}, including classical shadows~\cite{paini2019approximate,huang2020predicting,huang2021demonstrating,vankirk2022hardware,hadfield2022measurements,akhtar2023scalable,hu2023classical,kunjummen2023shadow,gandhari2024precision,levy2024classical,zhu2024connection,cai2024biased,hu2025demonstration}, which bypass full tomography and often allow efficient estimation of many properties directly from measurement data. The latter are especially popular in experiments~\cite{brydges2019probing,joshi2020quantum,joshi2024observing,zhang2021experimental,struchalin2021experimental,satzinger2021realizing,vitale2024robust,hu2025demonstration,dong2025measuring,andersen2025thermalization,votto2025learningmixedquantumstates}, since many variants can be realized on current devices using only simple operations and computational-basis measurements.

\begin{figure*}[t]
\centering
\includegraphics[width=2\columnwidth, trim = 0 80 0 100]{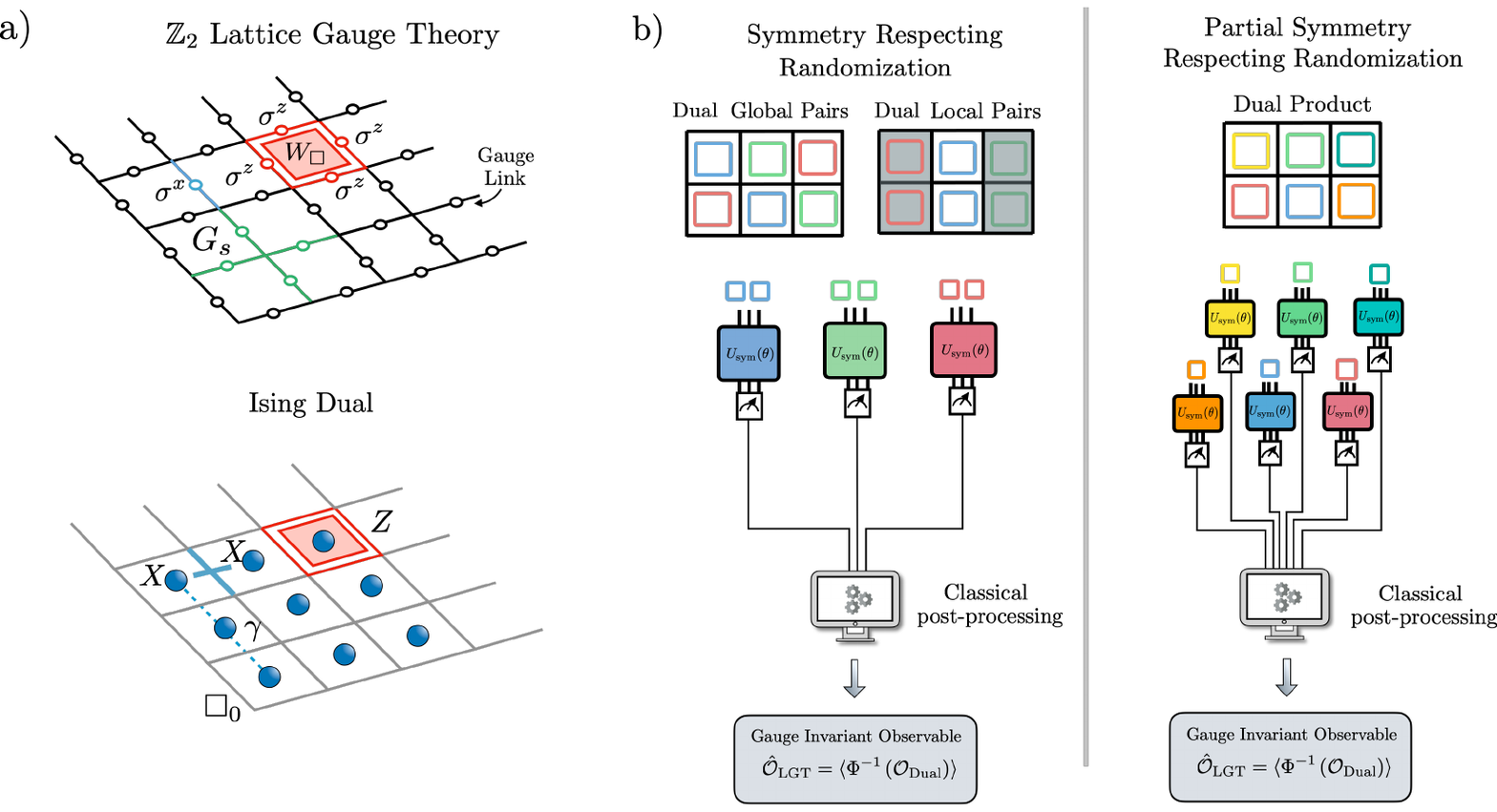}
\caption{\emph{Sample-efficient measurements of gauge-invariant observables in $\mathbb{Z}_2$ LGT.}
(a) Illustration of  $\mathbb{Z}_2$ LGT in $(2+1)$D and its Ising dual. For the LGT qubits reside on links $l$ of the lattice and Gauss law constraints $G_s$ are associated with sites $s$. The Hamiltonian consists of plaquette (magnetic) terms $W_\square=\prod_{l\in\square}\sigma^z_l$ and on-site (electric) terms $\sigma^x_l$. In the dual description as a $(2+1)$D Ising model the qubit degrees of freedom are associated with the plaquette. There is a one-to-one correspondence between operators and states within the physical Hilbert space of the LGT, independent of boundary conditions: the magnetic terms map to Pauli-$Z$ operators on the Ising qubits and the electric terms map to two-body Pauli-X terms acting on the qubits across the relevant link. In the inverse map from the Ising model to the LGT, single $X$ operators on the Ising side of the duality map to a string of $\sigma^x$ operators along some path $\gamma$ to an arbitrary reference plaquette $\square_0$. For PBC, the dual Ising model obeys a global parity constraint and such single-body $X$ operators are unphysical. (b) We introduce three symmetry-aware random measurement protocols—Global and Local Dual Pairs and Dual Product—designed to estimate gauge-invariant observables efficiently, leveraging prior knowledge that expectation values of gauge-variant operators vanish. For Global Dual Pairs, random pairs of Ising qubits are chosen and random two-body symmetry-respecting unitaries associated with these degrees of freedom are mapped back to the LGT side of the duality as the randomizing operations prior to measurement. Local Dual Pairs is similar but leverages a promise of geometric locality in the observables of interest to limit the choice of pairs to local patches. Finally, the Dual Product protocol is constructed from the standard Product Protocol applied to the dual Ising model. This protocol slightly breaks symmetry when we have periodic boundary conditions (PBC) since such operations are not parity-respecting. A trade-off exists between sample complexity and circuit depth. Symmetry-aware protocols offer exponential improvements in sampling efficiency at the cost of increased circuit depth over the standard symmetry-ignorant Product Protocol.
\label{fig:z2LGT}}
\end{figure*}

Randomized measurement protocols are designed to be general and state-agnostic, following the philosophy of “measure first, ask questions later”~\cite{elben2023randomized}. This flexibility, however, can increase the learning cost unnecessarily. In practice, prior knowledge about the structure of unknown states or relevant observables can guide measurement pre-processing~\cite{rath2021importance, huang2021efficient}. Key examples include symmetries—such as particle number or permutation symmetry—arising from the target state, the observables, or hardware constraints on measurement operations~\cite{elben2018renyi,vankirk2022hardware,hao2022classical,bringewatt2024randomized,hearth2024efficient,sauvage2024classical,arvind2025quantum}. A common feature is that exploiting prior knowledge reduces the measurement cost but can increase circuit complexity. 

Systems with \textit{local} symmetries, such as Lattice Gauge Theories (LGTs), exhibit distinctive Hilbert space structures and state properties. Quantum simulation of LGTs is a key application for quantum computers and simulators~\cite{halimeh2025quantum}, owing to their central role in describing high-energy and nuclear physics~\cite{Banuls:2019bmf,Klco:2021lap,bauer2023quantum,Bauer:2023qgm,di2023quantum}, condensed matter systems~\cite{fradkin2013field,kleinert1989gauge,wen1990topological,levin2005string}, fermion-to-qubit mappings~\cite{chen2018exact,chen2020exact,chen2023equivalence}, and because of connections to quantum error correction~\cite{sarma2006topological,nayak2008non,lahtinen2017short}.
We will show that incorporating prior knowledge when designing randomized measurement schemes for LGTs leads to especially pronounced trade-offs between sample complexity and circuit complexity. The common eigenspace of local symmetry constraints is exponentially smaller than the full Hilbert space used in the simulation (though still exponentially large in system size) and this can enable large, even exponential, improvements in sample complexity. However, such schemes require more complex circuit implementations, and analytical tasks—particularly the analysis and inversion of the associated shadow channels—become challenging.

In this manuscript, we introduce three protocols for sample-efficient measurement of gauge-invariant observables in $\mathbb{Z}_2$ LGT in (2+1)D, for which a dual Ising formulation allows rigorous analysis. We analyze the trade-offs between sampling and circuit complexity, provide an analytic inversion of the quantum channel underlying the shadow protocol, and establish rigorous performance guarantees. The three variants are:
\begin{itemize}
    \item[a.] \textbf{Dual Pairs Protocols}. These are inspired by the “All-Pairs” approach of Ref.~\cite{hearth2024efficient}, where shot-wise random connectivity between qubit pairs is used to implement particle-number-preserving randomization. Our protocols additionally leverage the LGT–Ising duality: most components are analyzed `virtually' via the duality, while physical operations are carried out on the LGT state realized on the device. Unlike Ref.~\cite{hearth2024efficient}, where particle-number-preserving randomization was introduced primarily for hardware reasons, here it directly reduces sampling cost. We develop two variants in Section~\ref{s:protocol}:
    \begin{itemize}
        \item[i.] The \textbf{Global Dual Pairs Protocol}, applicable to arbitrary gauge-invariant observables, exponentially improves sampling efficiency over standard protocols.

        \item[ii.] The \textbf{Local Dual Pairs Protocol} is tailored for geometrically local gauge-invariant observables, yields further improvements in sampling efficiency and reduces both sampling and circuit complexity compared to the Global Dual Pairs Protocol.
    \end{itemize}
    \item[b.] The \textbf{Dual Product Protocol} exploits the Ising–LGT duality without relying on an all-pairs structure. It achieves the best sampling efficiency among our protocols but is also the most demanding in terms of circuit resources, requiring depth $\mathcal{O}(V^2)$ (with $V$ the system volume) and an additional ancilla qubit for periodic boundary conditions; it is discussed in Section~\ref{s:dual-product-protocol}.
\end{itemize}
All protocols are benchmarked against the `standard' classical shadow protocol,  the \textbf{Product Protocol}, which applies single-qubit Clifford randomization~\cite{huang2020predicting}, and which uses no prior information about the state or the observables to be predicted.

Our work also complements recent studies of \emph{deep-randomization} circuits, i.e., approximate symmetric unitary k-designs, for LGTs~\cite{bringewatt2024randomized}. In contrast to these deep-randomization results, which primarily target the estimation global properties such as entanglement entropies, the shallow protocols developed here are tailored for efficient estimation of gauge-invariant observables, and they are substantially simpler to implement in near-term quantum devices. Moreover, while Ref.~\cite{bringewatt2024randomized} is numerical, our work provides rigorous asymptotic performance guarantees.

The paper is organized as follows:
In \cref{s:duality}, we introduce the $\mathbb{Z}_2$ LGT and its duality to the Ising model. Section~\ref{s:protocol} presents the two variants of the Dual Pairs Protocol---the Global Dual Pairs and Local Dual Pairs protocols. In Section~\ref{s:dual-product-protocol}, we describe the Dual Product Protocol. Numerical demonstrations of all protocols are given in \cref{s:example}. Finally, in \cref{s:discussion}, we discuss the implications of our results and outline future directions, including extensions to other LGTs, potentially beyond the Abelian case.

\section{Lattice Gauge Theory}\label{s:duality}
Lattice Gauge Theories (LGTs), are models in which the local degrees of freedom, on links (i.e.\ edges of the lattice graph), are associated with elements of a group, subject to a set of local symmetry operators imposing constraints on the Hilbert space that involves their neighbors. These operators, Gauss laws, may not mutually commute (they do for LGTs based on Abelian groups), but  commute with the Hamiltonian ensuring invariance of eigenstates under gauge transformations, see e.g., Refs.~\cite{kogut1975hamiltonian,kogut1979introduction,davoudi2025tasi} for reviews.

A simple example is the $\mathbb{Z}_2$ LGT~\cite{wegner1971duality,horn1979hamiltonian} in 2+1 spacetime dimensions. The underlying Abelian group $\mathbb{Z}_2$ has two elements, corresponding to local states encoded in spin-$\tfrac{1}{2}$ degrees of freedom (qubits) residing on the links of a rectangular $N_x \times N_y$ lattice; see \Fig{fig:z2LGT}(a). The Hamiltonian is given by
\begin{equation}\label{eq:H}
H=-\sum_\square W_\square-g\sum_l\sigma^x_l,
\end{equation}
where $W_\square := \prod_{l \in \square} \sigma^z_l$ is the plaquette operator and $\sigma^{x,z}_l$ are Pauli operators acting on link $l$. Gauss law operators  are defined at each site $s$ (node of the lattice graph) as
\begin{equation}
G_s = \prod_{l \in s} \sigma^x_l \,,
\end{equation}
where $l\in s$ denotes links emanating from a site. These operators commute with the Hamiltonian, $[H, G_s] = 0$, and thus label superselection sectors; that with $G_s | \psi\rangle = | \psi\rangle$ $\forall s$ is usually of interest (called the ``physical'' subspace). Gauge-invariant operators are invariant under local symmetry transformations 
\begin{align}
U O_{S} U^\dagger = O_{S},\, \qquad
U := \prod_{s \in S} (G_s)^{\alpha(s)}\,,
\end{align}
and their expectation value non-zero in a physical state. Here, 
 $\alpha(s)$ is an integer for cyclic groups such as $\mathbb{Z}_2$ ($\alpha(s)\in \{0,1\}$ for ${\mathbb{Z}_2}$), and $S$ denotes the set of sites (or links between sites) on which $O_{S}$ has support. For LGTs with continuous groups, $G_s$ are replaced by exponentiated Gauss law \textit{generators}, $G_s \rightarrow e^{iG_s}$ with real, continuous $\alpha(s)$; multiple non-commuting Gauss law operators exist per site in the non-Abelian case~\cite{kogut1975hamiltonian,kogut1979introduction,davoudi2025tasi}.

A lattice can feature different boundary conditions, which can lead to additional superselection sectors. For periodic boundary conditions (PBCs), discussed in the main text,  two additional operators $V_x \equiv \prod_{\ell \in \mathcal{P}_x}\sigma^x_\ell$, and $V_y \equiv \prod_{\ell \in \mathcal{P}_y}\sigma^x_\ell$ define superselection sectors; the periodic paths $\mathcal{P}_{x/y}$ are shown in the right panel of \cref{fig:z2LGT}(a) in blue and red, respectively. Other boundary conditions, such as fixed boundary conditions (FBC) are discussed in Appendix~\ref{a:global-FBC}.

The $\mathbb{Z}_2$ LGT is dual to an Ising model in the following sense:
Let $\mathcal{L}(\mathcal{H})$ denote the space of bounded linear operators acting on the (physical) LGT Hilbert space $\mathcal{H}_{\text{phys}}$. A fixed $(V_x,V_y)$ sector can be mapped exactly onto a dual Ising model~\cite{wegner1971duality,horn1979hamiltonian,mueller2022thermalization,hartse2024stabilizer}, using the  map,  
$\Phi:\mathcal{L}(\mathcal{H}_{\text{phys}})\rightarrow\mathcal{L}(\mathcal{H}_{\text{dual}})$,
with dual (qubit) d.o.f.s placed at the centers of plaquettes $\square$, 
\begin{align}\label{eq:duality-map}
\Phi(W_\square) = Z_\square\, , \qquad
\Phi(\sigma^x_l) = X_\square X_{\square+\hat{a}}\,,
\end{align}
where, w.o.l.g, $V_x=V_y=1$,  $X_\square$ and $Z_\square$ are Pauli operators, and $\square+\hat{a}$ are the plaquettes adjacent to link $l$.
The inverse map is 
\begin{align}\label{eq:inverse-duality-map}
\Phi^{-1}(Z_\square) = W_\square\, , \qquad
\Phi^{-1}(X_\square) = \prod_{l\in\gamma_\square} \sigma^x_l\,.
\end{align}%
Here $\gamma_{\square}$ is a path from a reference plaquette $\square_r$ to $\square$, as depicted in the bottom left panel of \cref{fig:z2LGT}(a).
The choice of $\square_r$ is arbitrary and no observable depends on it.
The corresponding dual Hamiltonian is
\begin{align}\label{eq:dualH}
H &= -\sum_\square Z_\square 
-g\sum_{\square} (X_\square X_{\square-\hat{x}}+X_\square X_{\square-\hat{y}}),
\end{align}
a (2+1)D transverse field Ising model with PBC;  $\square-\hat{x}$ ($\square-\hat{y}$) are the plaquette to the left (below) of $\square$. For PBC (but not for other boundary conditions), the product of plaquette operators satisfies the operator identity $\prod_\square W_\square = 1$. This maps, on the Ising side, to a  parity constraint $\prod_\square Z_\square = 1$ required for the duality to hold.
A counting of degrees of freedom confirms the exactness of the duality: the LGT has $2N_x N_y$ qubits (one per link), subject to $N_x N_y - 1$ independent Gauss law constraints. Including the decomposition into sectors labeled by $V_x$ and $V_y$, this results in $N_x N_y - 1$ physical degrees of freedom per sector—precisely matching the number of d.o.f.s in the parity-constrained Ising model.
Thus, the physical Hilbert space within a given $V_x,V_y$ sector is of dimension $2^{N_xN_y-1}$ whereas the full Hilbert space is exponentially larger with dimension $2^{2N_xN_y}$.
The duality, which can be generalized to arbitrary boundary conditions, will allow us to design several symmetry-respecting random measurement protocols, and  to derive explicit performance guarantees, in the next two sections.
\begin{figure*}[t]
\centering\includegraphics[width=\textwidth,trim ={0 85 0 80}]{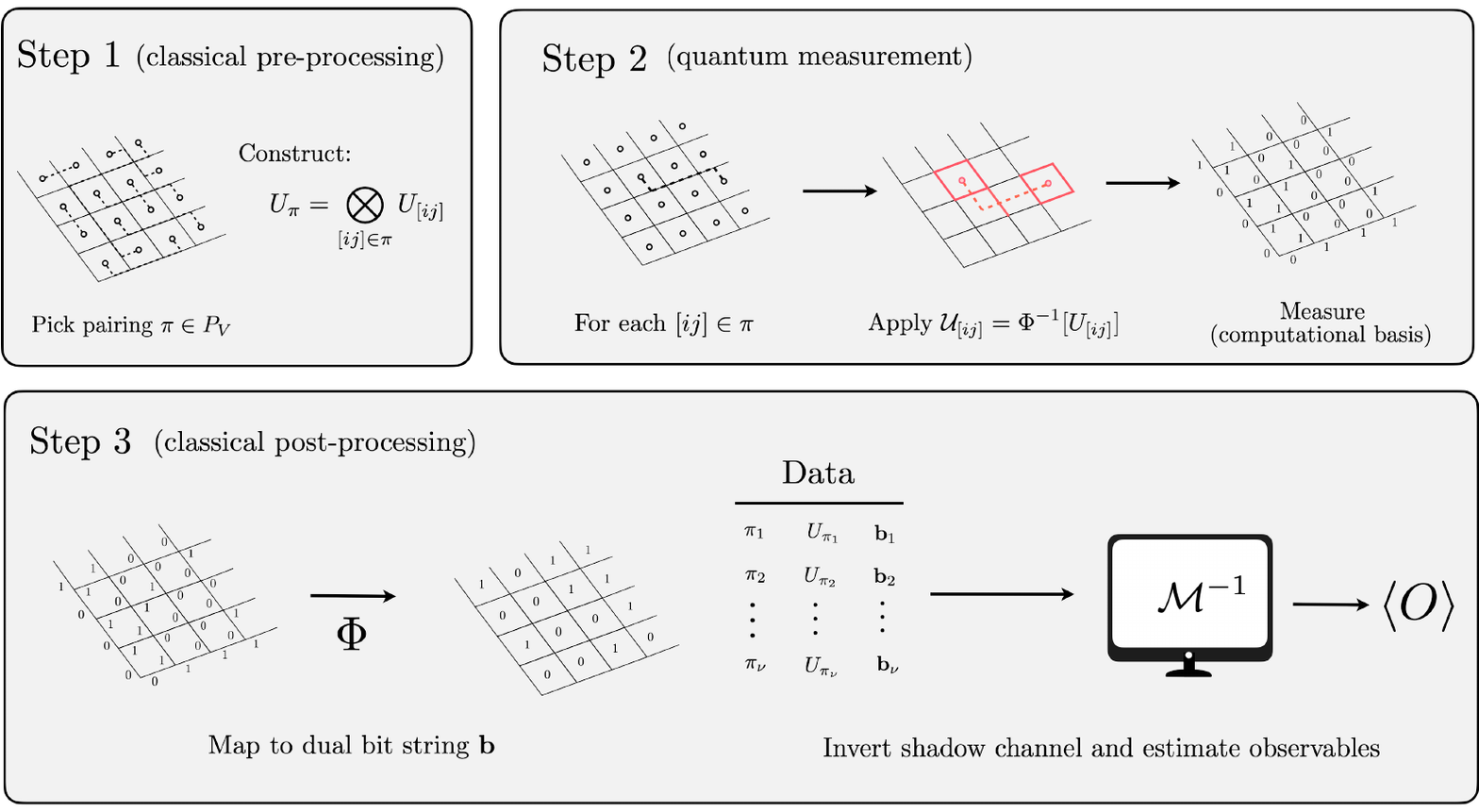}
\caption{\emph{Schematic overview of the Global Dual Pairs protocol with periodic boundary conditions (PBC).} \textbf{Step 1.} A pairing, $\pi$, of dual lattice sites is chosen uniformly at random.  For each pair, a random unitary $U_{[ij]}$ is constructed in the Ising formulation. \textbf{Step 2.} For each pair $[ij]\in\pi$, the dual unitary $\mathcal{U}_{[ij]}=\Phi^{-1}[U_{[ij]}]$ is implemented on the LGT side of the duality and a computational basis measurement is performed. \textbf{Step 3.} The output bit string is mapped back to the dual Ising theory, where efficient shadow channel inversion is performed to estimate expectation values of gauge-invariant observables $O$.}
\label{fig:global-all-pairs}
\end{figure*}

\section{Dual Pairs Protocols}\label{s:protocol}
Here, we present two protocols—the Global Dual Pairs and Local Dual Pairs schemes—for estimating gauge-invariant observables. Both protocols improve sample complexity compared to the symmetry-ignorant standard Product Protocol~\cite{huang2020predicting}, which applies local Clifford randomization, here, to the $2N_xN_y$ qubits associated with the links of the $\Ztwo$ LGT. The advantage arises because the Product Protocol randomizes the exponentially larger full Hilbert space, whereas our protocols randomize only over the physical subspace. The Global Dual Pairs protocol is the most general, requiring only that the target observables be gauge-invariant, and achieves an exponential reduction in sampling complexity relative to the Product Protocol. The Local Dual Pairs protocol, designed for geometrically local gauge-invariant observables, provides additional improvements in both sampling and circuit complexity.

\subsection{Global Dual Pairs Protocol}\label{s:global-protocol}
\subsubsection{Overview}
The Global Dual Pairs protocol for estimating gauge-invariant observables, without additional restrictions, is illustrated schematically in~\cref{fig:global-all-pairs}. For simplicity of presentation, we assume that both $N_x$ and $N_y$ are even, although this is not a requirement. The system realized in experiment, where measurements are performed, is the LGT; the dual Ising model serves purely as a conceptual tool for constructing the protocol and a computational tool for the classical pre- and post-processing. It is applied to a physical input state in the LGT, i.e., a simultaneous $+1$ eigenstate of all Gauss law operators. The protocol proceeds as follows:

\paragraph*{Step 1 (Classical Pre-processing):} Consider the dual Ising model consisting of $V = N_x \cdot N_y$ spins. Select, uniformly at random, a pairing $\pi$ that partitions all spins into disjoint pairs. For each pair $[ij] \in \pi$, choose a random two-qubit unitary $U_{[ij]}$. For systems with PBCs, a technical, but non-essential, requirement is that $U_{[ij]}$ respects the $\mathbb{Z}_2$ parity symmetry, as detailed below. No such constraint applies e.g. for FBCs. This Ising-side random unitary factorizes as
\begin{equation}\label{eq:ising-unitary}
U_\pi=\bigotimes_{[ij]\in\pi} U_{[ij]}.
\end{equation}

\paragraph*{Step 2 (Quantum Measurement):} Perform the random measurement on the quantum state on the $\Ztwo$ LGT side of the duality associated with the dual unitary corresponding to \cref{eq:ising-unitary}:
\begin{equation}
\mathcal{U}_\pi:=\Phi^{-1}(U_\pi).
\end{equation}
A circuit decomposition of $\mathcal{U}_\pi$ can be computed from $U_\pi$ with classical time complexity $\mathcal{O}(V)$  by applying \cref{eq:inverse-duality-map} to each of the Pauli rotation gates in the circuit decomposition of $U_\pi$. Following this rotation, perform computational basis measurements on all LGT qubits to obtain a bit string $\vec{s}\in\{0,1\}^{2 V}$.

\paragraph*{Step 3 (Classical Post-processing):} Map the bit string $\vec{s}$ to an associated Ising dual bit string $\vec{b}\in\{0,1\}^{V}$ for classical post-processing on the Ising side of the duality. In particular, 
\begin{equation}\label{eq:bit-string-map}
\ket{\vec{b}}\bra{\vec{b}}:=\Phi\left(\ket{\vec{s}}\bra{\vec{s}}\right).
\end{equation}
Mapping from a bit string $\vec{s}$ to a bit string $\vec{b}$ reduces to determining the parity around each plaquette and can be performed with classical time complexity $\mathcal{O}(V)$. Details can be found in \cref{s:bitstring-mapping}.

Finally, invert the shadow channel $\mathcal{M}$ defined on the Ising side of the duality as
\begin{equation}\label{eq:channel}
\mathcal{M}\circ\Phi\left(\rho\right):=\underset{{U,\vec{b}}}{\mathbb{E}}\left[U^\dagger\ket{\vec{b}}\bra{\vec{b}}U\right],
\end{equation}
where $\circ$ denotes the composition of linear maps and $U$ is the Ising-side unitary generated in Step 1. Here, $\rho$ is the state of the system as on the LGT side of the duality and $\Phi\left(\rho\right)$ is the associated state in the dual picture. This channel can be easily inverted to obtain an Ising-side classical shadow 
\begin{equation}\label{eq:shadow}
\widehat{\Phi\left(\rho\right)}:=\mathcal{M}^{-1}\left(U^\dagger \ket{\vec{b}}\bra{\vec{b}}U\right),
\end{equation}
associated with each sampled $U, \vec{b}$. The
details are described below.  Using these shadows, expectation values of local observables can be computed once the observables of interest have also been mapped to the dual picture, again with a classical cost $\in\mathcal{O}(V)$. 

At this point, it is important to remind the reader that the experiment, including the states, unitary operations, and measurements,is implemented on  LGT states (in Step 2), while the Ising duality is only a conceptual and (classical) numerical tool for defining the  unitaries and efficiently inverting the associated channel.

\subsubsection{Random Symmetry-Respecting Unitaries}\label{ss:implementing-random-unitaries}
\begin{figure}[t]
\includegraphics[width=0.9\columnwidth,trim ={0 175 0 150}]{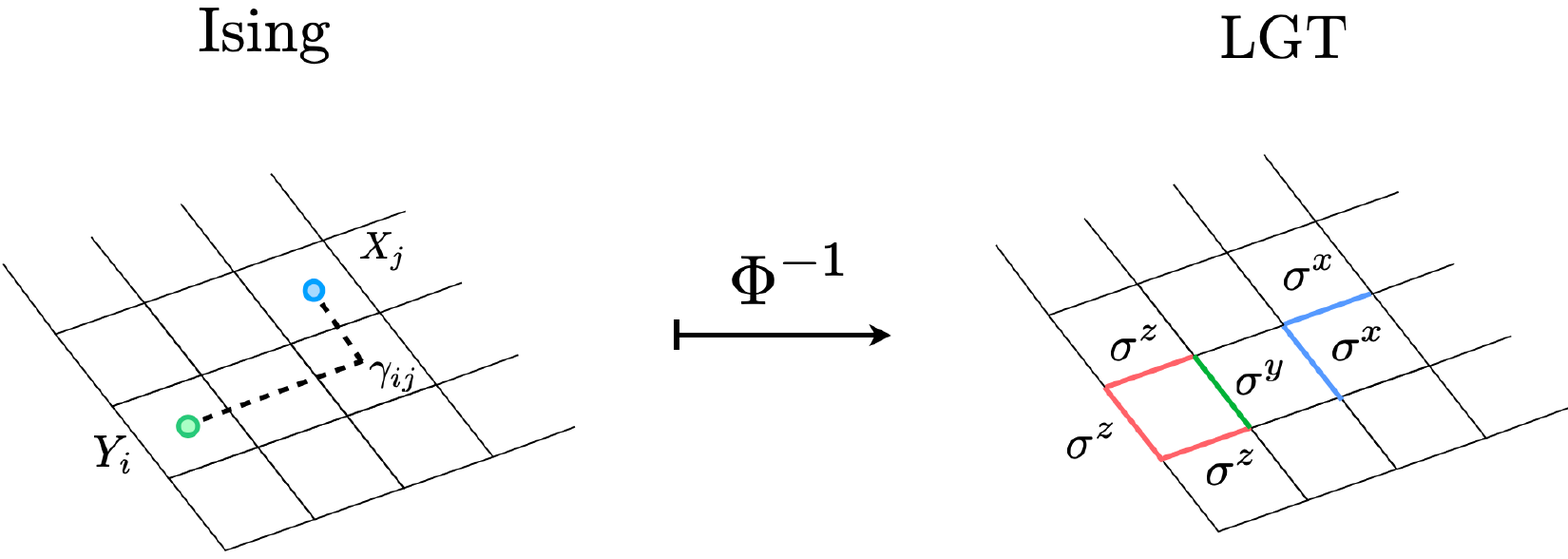}
\caption{\emph{LGT–Ising duality for mapping a (parity-respecting) entangling operation $\Phi^{-1}(Y_i X_j)$.}  While this operation is two-body on the Ising side (top panels), it corresponds to an extended operation in the LGT (bottom panels) acting between plaquettes $i$ and $j$ along a path $\gamma_{ij}$, which is arbitrary except for its fixed endpoints.\label{fig:circuits}}
\end{figure}

In this section, we detail the structure and cost of implementing the random symmetry-respecting unitaries $\mathcal{U}_\pi$ constructed in the Ising picture in \textbf{Step 1} and implemented on the LGT states in \textbf{Step 2}. For systems with PBCs, we choose each local unitary $U_{[ij]}$  to respect parity symmetry. In the absence of a symmetry constraint, $U_{[ij]}$  instead is simply chosen as a generic two-qubit Haar-random unitary. Specifically, for the case with parity symmetry, we consider a decomposition,
\begin{equation}\label{eq:Uij}
    U_{[ij]}=U_{\mathrm{odd}}\oplus U_{\mathrm{even}},
    \end{equation}
    where $U_{\mathrm{odd}}, U_{\mathrm{even}}$ are Haar random $2\times 2$ unitaries independently drawn from a circular unitary ensemble (CUE) acting on the relevant parity sector. Such unitaries can be parameterized, up to an irrelevant global phase, as~\cite{brydges2019probing} 
    \begin{align}\label{eq:1qubbitrotation}
     U_{\substack{\rm odd/ \\ \rm even}}=  \begin{pmatrix}
     e^{i(\alpha+\gamma)}\cos(\beta) & e^{-i(\alpha-\gamma)}\sin(\beta) \\
    - e^{i(\alpha-\gamma)}\sin(\beta) & e^{-i(\alpha+\gamma)}\cos(\beta) 
    \end{pmatrix}\,
    \end{align}
    for appropriate choices of $\alpha$, $\beta$, and $\gamma$, with circuit realization 
    \begin{equation}\label{eq:dual-circuit}
    \begin{adjustbox}{width=0.45\textwidth}
    \begin{quantikz}[transparent, row sep=0.12cm]
    \lstick{$i$}\qw & \gate{e^{\;i\frac{\alpha}{2} Z} \;} \qw & \gate[2]{U_-(\beta)}\qw & \gate{e^{\;i\frac{\gamma}{2} Z}\;} & \gate{e^{\;i\frac{\alpha}{2} Z} \;} \qw & \gate[2]{U_+(\beta)}\qw & \gate{e^{\;i\frac{\gamma}{2} Z}\;}& \qw \\
    \lstick{$j$} \qw & \gate{e^{-i\frac{\alpha}{2} Z}} \qw & \qw & \gate{e^{-i\frac{\gamma}{2} Z}} & \gate{e^{-i\frac{\alpha}{2} Z}} \qw & \qw & \gate{e^{-i\frac{\gamma}{2} Z}}&\qw ,
    \end{quantikz}
    \end{adjustbox}
    \end{equation}
    where $U_{\pm}(\beta)\equiv \exp\{i\frac{\beta}{2}( Y_i X_j \pm X_i Y_j)\} $ correspond to the even/odd parity sectors. (We work in the even sector.)  
To translate this circuit to the LGT side of the duality, we use
\begin{align}
\Phi^{-1}(Z_\square)&=W_\square, \qquad \square\in\{i,j\},
\end{align}
and
\begin{align}\label{eq:YX-terms}
&\Phi^{-1}(Y_iX_j\pm X_iY_j)=\nonumber\\
&\qquad i\prod_{l\in\gamma_{ij}} \sigma^x_l\prod_{l'\in \square_j} \sigma^z_{l'}
\pm i\prod_{l\in\gamma_{ij}} \sigma^x_l\prod_{l'\in \square_i} \sigma^z_{l'},
\end{align}
where $\gamma_{ij}$ is some choice of path between plaquettes $i$ and $j$, as depicted in \cref{fig:circuits} for the $Y_iX_j$ term of this operator. 

\cref{eq:YX-terms} implies that implementing $\Phi^{-1}(U_\pm(\beta))$ requires a unitary rotation with Pauli weight that scales with the distance between the plaquettes $i,j$. As the Pauli weight of a  rotation directly determines the circuit depth of a decomposition into a standard gate set, this results in a higher circuit depth when compared to the standard Product protocol; for the Global Dual Pairs protocol it is $\mathcal{O}(V)$. 
Another consideration is the choice of paths $\gamma_{ij}$, which determine how the unitaries $\mathcal{U}_{[ij]}$ are implemented in parallel. In practice, a reasonable, though not necessarily optimal, solution to this path selection problem can be obtained using a greedy algorithm~\cite{blackdictionary} that identifies non-intersecting paths for each pair $[ij] \in \pi$. 

As an upper bound, without parallelization, the circuit depth scales as $\mathcal{O}(V^2)$: one factor of $V$ arises from the number of such pairs, and the second from the decomposition of each high-weight Pauli rotation into elementary gates. In \cref{s:compare-pauli} we shall see that the large circuit depth for the Global Dual Pairs protocol compared to the Product protocol is counteracted by a large reduction in sample complexity.

\subsubsection{Mapping of Measurement Outcomes}\label{s:bitstring-mapping}
In this subsection, we discuss the first part of \textbf{Step 3} of the Global Dual Pairs protocol where we map measurement outcomes $\vec{s}\in\{0,1\}^{2V}$ to dual bit strings $\vec{b}\in\{0,1\}^V$ as described in \cref{eq:bit-string-map}.  The mapping is relatively simple: the parity of the measurement outcomes around a given plaquette corresponds to the measurement outcome that would result if we had directly measured $Z_\square$ on that same plaquette in the dual Ising theory. Thus, computing the parity of $\vec{s}$ for each plaquette allows us to determine the bit string $\vec{b}$ that results from a computational basis measurement in the dual picture. This process requires classical time complexity $\mathcal{O}(V)$. After this mapping, we treat the inversion of the channel purely as a classical shadows scheme in the dual picture, as elaborated further in the following section.

\subsubsection{Inverting the Channel}\label{s:channel-inversion}
In this subsection, we discuss the second part of \textbf{Step 3} of the Global Dual Pairs protocol—--the inversion of the channel defined in \cref{eq:channel}. We focus on the case with PBCs, where the Ising dual retains a global parity constraint. Our approach to analyzing the channel inversion adapts techniques originally developed in Ref.~\cite{hearth2024efficient} for systems with particle number conservation. Other boundary conditions, where no such symmetry remains on the Ising side, lead to somewhat different (but simpler) inversion procedures and are discussed in \cref{a:global-FBC}.

Let $P_V$ denote the set of all pairings of the $N_xN_y=V$ sites of the dual lattice (centers of plaquettes in the original lattice). 
Assuming that $N_x$ and $N_y$ are both even, $V$ is also even so that all sites can be paired up. The number of pairings is $|P_V|:=(V-1)!!$. Formulated on the Ising side of the duality, the Global Dual Pairs channel is given as an average over channels defined for each of the possible pairings $\pi\in P_V$ (as each pairing is selected with uniform probability):
\begin{align}
\mathcal{M}(\cdot)&=\frac{1}{|P_V|} \sum_{\pi\in P_V} \mathcal{M}_\pi(\cdot).
\end{align}
Then, $\mathcal{M}_\pi(\rho)$ decomposes into a product over channels acting non-trivially on each pair $[ij]\in\pi$ yielding
\begin{align}
\mathcal{M}(\cdot)&= \frac{1}{|P_V|} \sum_{\pi\in P_V} \prod_{[ij]\in\pi} \mathcal{M}_{[ij]}(\cdot),
\end{align}
Finally, these pairwise channels act on the Ising-side state $\Phi(\rho)$ as
\begin{equation}
\mathcal{M}_{[ij]}\circ\Phi\left(\rho\right)=\underset{{U_{[ij]},b_i, b_j}}{\mathbb{E}}\left[U^\dagger_{[ij]}\ket{b_ib_j}\bra{b_ib_j}U_{[ij]}\right],
\end{equation}
where $U_{[ij]}$ are two-qubit unitaries selected according to \cref{eq:Uij} and $b_i, b_j$ are the measurement outcomes mapped to the dual qubits $i,j$, as described in the previous section. 
\begin{figure*}[t]
\centering\includegraphics[width=\textwidth,trim ={0 85 0 80},clip]{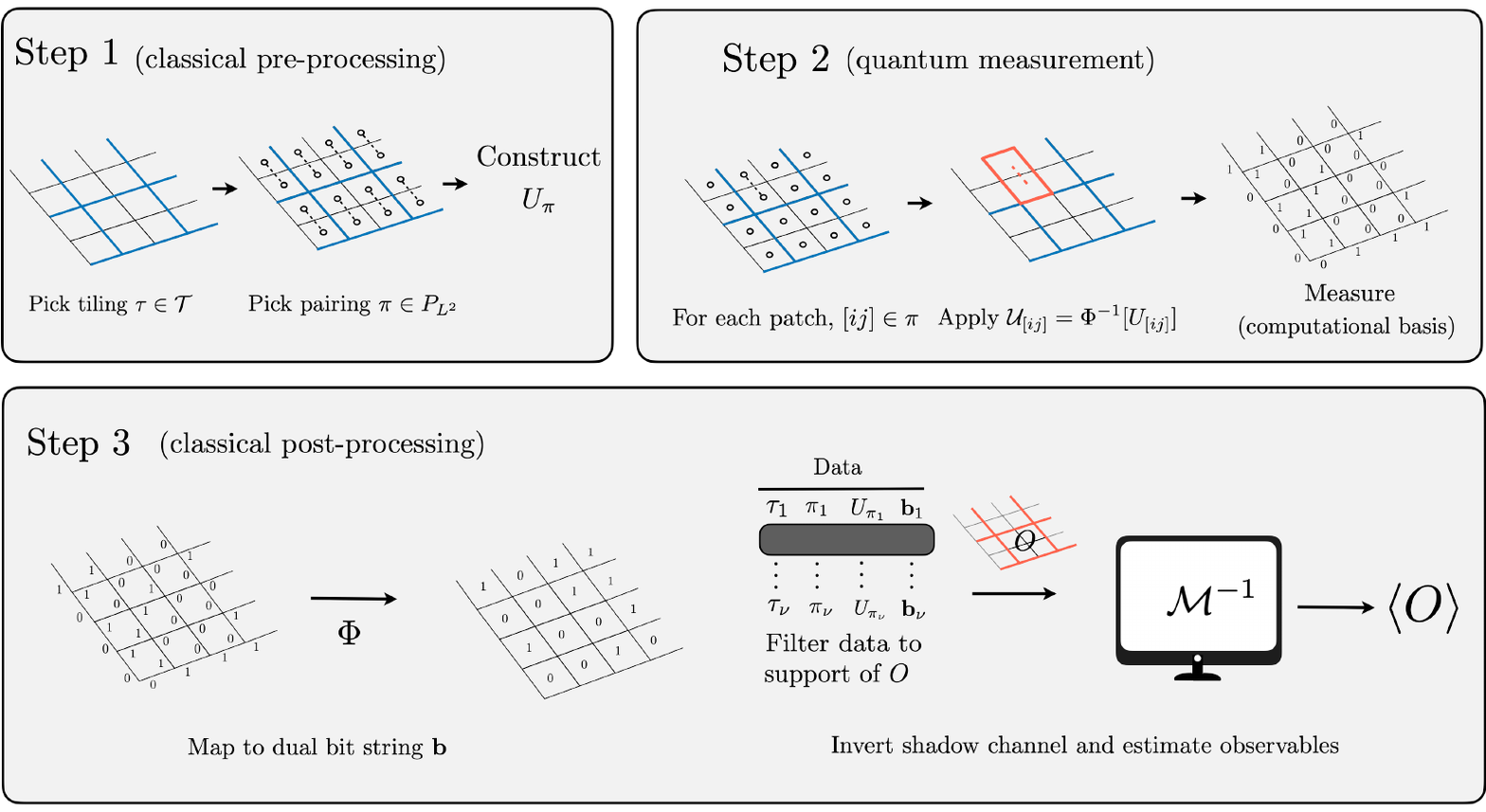}
\caption{\emph{Overview over the Local Dual Pairs protocol.} The Local Dual Pairs protocol is a variant of our scheme for geometrically local gauge-invariant observables that have support on a patch of at most $L\times L$. \textbf{Step 1} consists of picking a random tiling $\tau\in\mathcal{T}$ and, then, picking a random pairing repeated within each patch of the tiling in order to construct an associated random unitary $U_\pi$. \textbf{Step 2} consists of applying the associated unitary $\mathcal{U}_\pi$ on the LGT side of the duality, followed by measurement in the computational basis. \textbf{Step 3} is the post-processing step, which proceeds essentially identically to the Global Dual Pairs protocol restricted to a given patch of a tiling, where the data is filtered to a single choice of tiling according to the  support of a given target variable.}\label{fig:local-all-pairs}
\end{figure*}

Because of the parity symmetry, the image $\Phi\left(\rho\right)$ block diagonalizes into even and odd parity sectors as 
\begin{equation}
\Phi\left(\rho\right)=\Phi\left(\rho\right)_\mathrm{even}\oplus\Phi\left(\rho\right)_\mathrm{odd}.
\end{equation}
The same decomposition occurs for the channel $\mathcal{M}(\cdot)$. Thus, we can evaluate $\mathcal{M}_{[ij]}\circ\Phi(\rho)$ as a direct sum over the even and odd parity blocks. Using that $U_\mathrm{odd}, U_\mathrm{even}$ are sampled from a 2-design, the combined action of channel and isomorphism on the state $\rho$ is~\cite{mele2024introduction}
\begin{align}\label{eq:two-design-action}
\mathcal{M}_{[ij]}\circ\Phi\left(\rho\right)=\bigoplus_{p\in\mathrm{even}, \mathrm{odd}} \frac{1}{3}\left(\Phi\left(\rho\right)_p+\mathrm{Tr}\big[\Phi\left(\rho\right)_p\big]\right).
\end{align}
Any gauge-invariant observable can be written on the Ising side of the duality in the basis of Pauli strings $S\in\{I,Z,X,Y\}^{\otimes V}$ with an even number of $X,Y$ operators.  Using \cref{eq:two-design-action}, we evaluate the action of $\mathcal{M}_{[ij]}$ on substrings $S_i S_j$,
\begin{align}\label{eq:channel-action}
&\mathcal{M}_{[ij]}[S_iS_j]= \nonumber\\
&\qquad\begin{cases}
S_iS_j, & S_iS_j\in\{II, ZZ\}\\
\frac{1}{3} S_iS_j,  &S_iS_j\in\{IZ, ZI\}\cup \{X,Y\}^2\\
0, & \text{otherwise}\,,
\end{cases}
\end{align}
indicating that $\mathcal{M}$ is diagonal, making inversion almost trivial: the non-trivial component of the inversion is all contained in the  Ising duality. In particular, for a Pauli string $S$,
\begin{align}\label{eq:channel-factorization}
\mathcal{M}\left(S\right) = \frac{f\alpha}{3^{w_{XY}/2}} S,
\end{align}
where $w_{XY} = w_X + w_Y$ is the total number of $X$ and $Y$ operators in the string $S$, $f$ is the fraction of pairings $\pi \in P_V$ for which $\mathcal{M}_\pi(S) \neq 0$, 
\begin{equation}\label{eq:f}
f=|P_{w_{XY}}|\frac{|P_{V-w_{XY}}|}{|P_{V}|},
\end{equation}
where $|P_{m}|$ is the number of pairings of $m$ objects, and $\alpha$ is an amplitude associated with permutations of the $Z$ and $I$ operators within $S$:
\begin{widetext}
\begin{align}\label{eq:aLocal Pairsha}
\alpha=\frac{1}{|P_{w_I+w_Z}|}\sum_{\substack{m=0 \text{ s.t.}\\ m \cong w_Z (\text{mod } 2)}}^{w_Z} \left(\frac{1}{3}\right)^m\underbrace{\binom{w_Z}{m}\binom{w_I}{m}m!}_{\text{num. of $IZ$ pairings}}\underbrace{|P_{w_Z-m}||P_{w_I-m}|}_{\text{num. of $ZZ$, $II$ pairings}},
\end{align}
\end{widetext}
where $w_Z$ and $w_I$ are the number of $Z$ and $I$ operators in the operator string $S$ respectively, and  $V=w_{XY}+w_Z+w_{I}$. The sum in \cref{eq:aLocal Pairsha} is over the number of pairs $m$ of $Z$ operators with $I$ operators, where the notation indicates that $m$ increments by two for each term in the sum, starting from either 0 or 1 depending on the parity of $w_Z$.  As seen in \cref{eq:channel-action} these pairings are the pairings that yield a non-trivial contribution to the amplitude. If $w_z=0$, then $\alpha =1$.

Upon inverting $\mathcal{M}$, the number of samples $\nu$ needed for predicting, with high probability, the expectation values of $M$ observables $\{O_j\}_{j=1}^M$ of interest up to an additive error $\epsilon$ can be bounded as~\cite{huang2020predicting},
\begin{equation}\label{eq:nubound}
\nu \in \mathcal{O}\left(\frac{\log M}{\epsilon^2}\max_j \mathrm{Var}(o_j)\right),
\end{equation}
where $o_j:=\mathrm{Tr}[\widehat{\Phi[\rho]} O]$, where $\widehat{\Phi[\rho]}$ is the (Ising side) shadow as given in \cref{eq:shadow}. The operator variance can be bounded as (see \cref{a:details-sample-complexity}):
\begin{align}\label{eq:var_gp}
\mathrm{Var}(o_j) &\leq 
\frac{3^{w_{XY}/2}}{f\alpha}
\norm{O_j}_\infty^2.
\end{align}
Thus, to reach constant error $\epsilon$ a number of samples scaling as $3^{w_{XY}/2}/(f\alpha)$ are needed. While the expressions for $f$ and $\alpha$ in \cref{eq:f,eq:aLocal Pairsha} are somewhat unwieldy, it is easy to determine whether the scaling is polynomial or exponential in system size. In particular, under the assumption that, asymptotically in system size, $w_I = \Theta(V)$ and $w_{XY}+w_Z=\mathcal{O}(1)$, we can use Stirling's approximation to show that
\begin{align}
f\alpha =\Theta\left(V^{(w_Z-k_\mathrm{dual})/2}\right).
\end{align}
Note that $w_Z\leq k_\mathrm{dual}$. Thus, for constant norm observables, a polynomial number of samples in the system size are needed to reach a constant error $\epsilon$.
The full costs of this protocol, including the $\mathcal{O}(V)$ costs of performing the mappings between the LGT and Ising  models and the $\mathrm{poly}(V)$ costs of computing the coefficients $f$ and $\alpha$, are summarized in \cref{tab:protocol-costs}.

\subsection{Local Dual Pairs Protocol}\label{ss:g-local-protocol}
In this section, we introduce a variant of the Global Dual Pairs protocol, referred to as Local Dual Pairs protocol, designed for the measurement of geometrically local observables rather than general $k$-local ones. This modification exploits the assumption that all observables of interest act non-trivially only on local patches of the lattice with constant size $L \times L$. Under this locality constraint, the sample complexity, classical cost of channel inversion, and the Pauli weight of the required quantum circuits are all significantly reduced.

For simplicity, we assume that $N_x$ and $N_y$ are divisible by $L$. We consider the set $\mathcal{T}$ of all distinct tilings of the lattice into adjacent $L\times L$ patches. As shifting a given tiling by $L$ units in either the $x$ or $y$ direction on the lattice recovers the same tiling, the number of tilings is $|\mathcal{T}|=L^2$. With this setup, the Local Dual Pairs protocol proceeds as follows:
\paragraph*{Step 1 (Classical Pre-processing):} Pick a random tiling $\tau\in\mathcal{T}$. Then, pick a pairing of $\pi\in P_{L^2}$ of sites in a single tile and consider that same pairing in each tile. The set of all such pairings $P_{L^2}$ per tiling $\tau \in \mathcal{T}$ has size $|P_{L^2}|=(L^2-1)!!$. As in the Global Dual Pairs protocol, construct a unitary $U_\pi$ associated with the pairing $\pi$.

\paragraph*{Step 2 (Quantum Measurement):} As in the Global Dual Pairs protocol, apply the unitary $\mathcal{U}_\pi=\Phi\left(U_\pi\right)$ on the $\Ztwo$ side of the duality and make a computational basis measurement.

\paragraph*{Step 3 (Classical Post-processing):} Again, as in the Global Dual Pairs protocol, map the output bit string to its corresponding bit string $\vec{b}$ in the Ising dual. Repeat the above sequence of steps a total of $\nu$ times. To estimate a geometrically local observable $O$, we filter the dataset to retain only those samples corresponding to a single $\tau \in \mathcal{T}$ for which $O$ is supported within one of its patches. 
This filtering is efficient as the number of tilings is $L^2$.
From this subset, we  restrict the corresponding dual output bit strings $\vec{b}$ and the unitaries $\bigotimes_{\text{patches}} U_\pi$ to the patch on which $O$ acts non-trivially. Using this restricted data, we perform the post-processing as in the Global Dual Pairs protocol on the local patch. For data restricted to this patch, the associated channel $\mathcal{M}$ coincides with that of the Global Dual Pairs protocol and can therefore be inverted in the same manner.

Given the filtered data of a single relevant tiling $\tau\in\mathcal{T}$, the fraction $f_\mathrm{local}$ of pairings that contribute in the Local Dual Pairs protocol
\begin{equation}
f_\mathrm{local}=|P_{w_{XY}}|\frac{|P_{L^2-w_{XY}}|}{|P_{L^2}|},
\end{equation}
and the coefficient $\alpha$ are calculated as in \cref{eq:aLocal Pairsha}, but with $w_I$ replaced by the number of identity operators in a given basis operator string $S$ within the relevant patch. 
\vspace{0.5em}

\noindent A diagram summarizing this protocol can be found in \cref{fig:local-all-pairs} and a summary of the cost  of the protocol can be found in \cref{tab:protocol-costs}.

\section{Dual Product Protocol}\label{s:dual-product-protocol}
At this stage, the reader may have noticed that the parity symmetry, and the associated “all-pairs” structure on the Ising side, are not essential for an efficient measurement scheme (indeed, no parity symmetry exists under other boundary conditions). The improvement in sampling complexity arises solely from exploiting the LGT–Ising duality which eliminates the exponential overhead between the gauge-invariant subspace and the full Hilbert space.

This suggests an alternative approach: (virtually) applying the standard Product Protocol on the Ising side of the duality, which we refer to as the Dual Product Protocol. The scheme is conceptually straightforward, apart from one important caveat: For PBC, the protocol mixes the even and odd parity sectors of the dual Ising theory, where only the even sector is dual to the LGT. On the LGT side, the parity constraint appears as an operator identity,
$\prod_\square W_\square =1$, rather than a symmetry: there is no component of the full LGT Hilbert space corresponding to the odd sector.

\begin{figure}
    \centering
\includegraphics[width=\columnwidth, trim={0 50 0 0}]{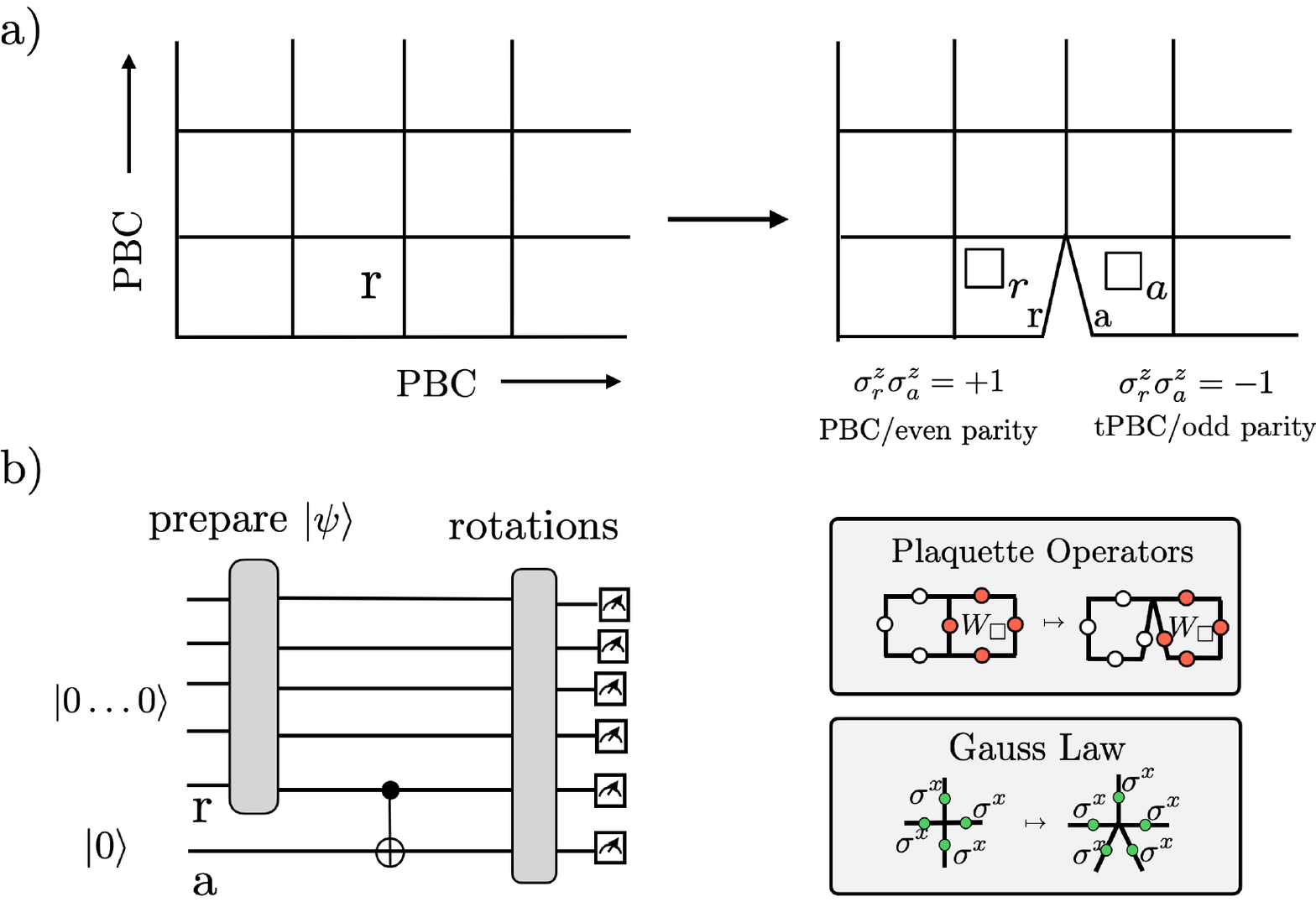}
    \caption{\textit{An ancilla enables mixing PBC (even parity) and tPBC (odd parity) sectors.} (a) Adding an ancilla $a$ to a reference link $r$ allows us to enforce either PBC or tPBC in the $\Ztwo$ LGT. (b) In the Dual Product protocol, the $z$-component of the state of the link $r$ is ``copied'' to the ancilla $a$ via a CNOT. The randomizing unitaries, chosen according to the usual Product Protocol, but applied to the Ising degrees of freedom, then mix between the PBC and tPBC sectors on the LGT side of the duality. In the right panel, we show how the plaquette and Gauss law operators change with the ancilla $a$: The plaquette operators no longer share a link, while the Gauss law operator is effectively extended by a fifth leg, $\sigma^x_r\rightarrow \sigma^x_r \sigma^x_a$.}
    \label{fig:tbc}
\end{figure}

To obtain a one-to-one correspondence between the full Ising theory (including both parity sectors) and an (extended) LGT, we introduce on the LGT side an ancilla qubit $a$ to control the lattice boundary condition, as shown in \Fig{fig:tbc}. Picking an \textit{arbitrary} reference plaquette $\square_r$ and selecting one of its four boundary links $r$, we designate that link as the control and the ancilla $a$, initially in the state $|0\rangle$, as the target of a ${\rm CNOT}_{r\rightarrow a}$ operation. As illustrated in \Fig{fig:tbc}, this operation effectively ``copies'' the magnetic ($z$-basis) state of link $r$ into the ancilla, while leaving the electric basis unchanged. The two plaquettes that originally shared link $r$ are thereby separated: one now contains $r$, and the other contains the ancilla $a$. The plaquette identity $\prod_{\square} W_{\square} = 1$ then becomes $\prod_{\square} W_{\square} = \sigma_a^z \sigma_r^z$.  Thus, we have effectively promoted the operator identity $\prod_\square W_\square =1$ to a symmetry $\prod_{\square} W_{\square}= \sigma_a^z \sigma_r^z$ with the original theory without the ancilla corresponding to the sector where $\sigma_a^z \sigma_r^z=1$.

Introducing the ancilla also affects electric operators (defined in the $x$-basis). In particular,
$
{\rm CNOT}_{r \rightarrow a} \sigma^x_r {\rm CNOT}_{r \rightarrow a} = \sigma^x_r \sigma^x_a,
$
so that every electric operator involving the reference link $r$—including Gauss-law and ribbon operators—is effectively modified as $\sigma^x_r \rightarrow \sigma^x_r \sigma^x_a$. For example, the two Gauss-law operators that originally shared $r$ now each become five-legged, see \Fig{fig:tbc}.

The states of the original $\mathbb{Z}_2$ LGT with periodic boundary conditions are mapped by the CNOT$_{r\rightarrow a}$ operation to the extended system including the ancilla where the original $z$-state of link $r$ is encoded in the two-qubit subspace ${ \ket{00}, \ket{11} }$ of the pair $(r, a)$. Crucially, however, the ancilla allows the boundary condition to be dynamically controlled by the state of $a$, enabling superpositions of PBC ($\prod_\square W_\square = 1$) and twisted PBC (tPBC, $\prod_\square W_\square = -1$). In this enlarged formulation of the LGT-Ising duality the \textit{entire} Ising space can now be represented on the LGT side: $\sigma_a^z \sigma_r^z = 1$ (PBC) corresponds to the even-parity sector, while $\sigma_a^z \sigma_r^z = -1$ (tPBC) corresponds to the odd-parity sector.\footnote{This is analogous, e.g., to how ancillas can induce disorder-free localization without explicitly breaking lattice symmetries~\cite{smith2017disorder,halimeh2024disorder}.} 

The Dual Product protocol now consists simply of single-qubit Clifford rotations on the Ising side, which mix the parity sectors. On the LGT side of the duality this mixes the $\prod_\square W_\square=1$ (PBC) and $\prod_\square W_\square=-1$ (tPBC) sectors. By virtue of the initial state being in the $\prod_\square W_\square=1$ (PBC) sector the shadow protocol that breaks this symmetry will still yield an estimate of $\prod_\square W_\square=1$, in expectation. 

In detail, the steps of the Dual Product Protocol are as follows:
\begin{figure*}[t]
\centering\includegraphics[width=\textwidth,trim ={0 100 0 80}]{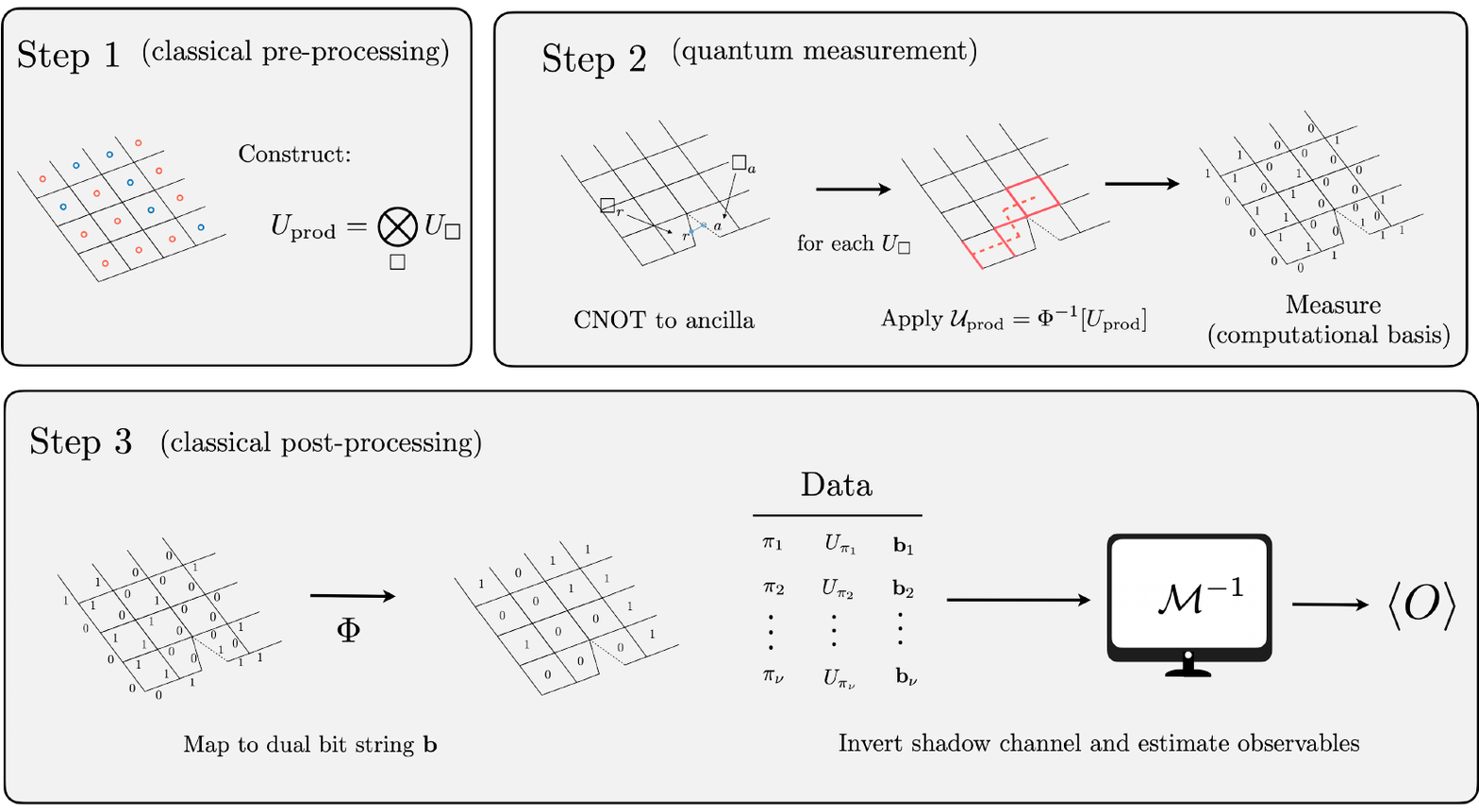}
\caption{\emph{Schematic overview of the Dual Product protocol with periodic boundary conditions (PBC).} \textbf{Step 1.} Randomly choose a unitary $U_\mathrm{prod}$ consisting of a single qubit Clifford gate on each Ising qubit. \textbf{Step 2.} The dual unitary $\mathcal{U}_\mathrm{prod}=\Phi[U_\mathrm{prod}]$ is implemented on the LGT side of the duality and a computational basis measurement is performed. \textbf{Step 3.} The output bit string is mapped back to the dual Ising theory, where efficient shadow channel inversion is performed to estimate expectation values of gauge-invariant observables $O$.}
\label{fig:dual-product}
\end{figure*}

\paragraph*{Step 1 (Classical Pre-processing):} 
Select a unitary operator $U_\mathrm{prod}$ on the dual Ising qubits of the form $U_\mathrm{prod}=\bigotimes_j U_j$ where each $U_j$ is a random single qubit Clifford gate acting on site $j$ (i.e. Hadamard or $S$ gates, which may be generated by single qubit $X$ and $Z$ rotations). 

\paragraph*{Step 2 (Quantum Measurement):} 
Introduce the ancilla $a$ in the LGT state and apply ${\rm CNOT}_{r \rightarrow a}$, effectively “copying” the $\sigma^z$-basis value of an arbitrary link $r$ into $a$,  while also transforming $\sigma^x_r \rightarrow \sigma^x_r \sigma^x_a$. The original state is preserved, now embedded in a slightly larger Hilbert space. 
Then, use $\mathcal{U}_\mathrm{prod} = \Phi^{-1}(U_\mathrm{prod})$ to apply operations on the LGT state: the generalized duality is understood to map two neighboring $Z$ operators on the Ising side to separate plaquettes on the LGT side—one containing $r$ and the other containing $a$. Simultaneously, every occurrence of $\sigma^x_r$ is replaced by $\sigma^x_r \sigma^x_a$ when mapping electric operations (i.e., those involving $X$ in the Ising dual). These operations mix the parity-even and parity-odd sectors of the Ising dual and create superpositions between PBC and tPBC in the LGT (see \cref{fig:dual-product} for an illustration). Following this we perform computational basis measurements to obtain a bit string $\vec{s}$.

\paragraph*{Step 3 (Classical Post-processing):} Map the bit string $\vec{s}$ to the corresponding Ising dual bit string $\vec{b}$ for channel inversion. Compared to the previously discussed protocols, $\vec{s}$ is now one bit longer due to the ancilla, but the mapping follows \cref{eq:bit-string-map}: the parity around each plaquette is determined, with the key difference that plaquette $\square_r$ no longer shares a link with its neighbor. As a result, the parities of the bit strings associated with each plaquette are now completely independent. Once $\vec{b}$ is obtained, post-processing reduces to the standard Product protocol applied to the Ising degrees of freedom, with a straightforward shadow channel inversion as outlined in~\cite{huang2020predicting}.

\vspace{0.5em}

We emphasize once more that the protocol effectively \textit{estimates} the LGT (magnetic) boundary condition (i.e., the parity on the Ising side), thus effectively discarding previously known information, with periodic boundary conditions appearing only in expectation.
The protocol can estimate \textit{arbitrary gauge invariant observables} for which its sample complexity  can be better than the Global Dual Pairs protocol:  In particular,
\begin{equation}
\mathrm{Var}(o_j) \leq 4^{k_\mathrm{dual}} \norm{O_j}_\infty^2,
\end{equation}
where $k_\mathrm{dual}$ is the locality of the operator of interest in the dual theory, and  $\norm{O}_\infty$ is evaluated on the dual Ising Hilbert space~\cite{huang2020predicting}. Thus, via \cref{eq:nubound}, the sample cost is constant in system size for gauge invariant observables that are $k_\mathrm{dual}$-local on the Ising side of the duality, while for Global Dual Pairs the sample cost is polynomial in system size.\footnote{Here, we use locality in the sense that a $k$-local operator acts non-trivially on $k$ qubits (has Pauli weight $k$), in contrast to the notion of geometric locality leveraged in the Local Dual Pairs protocol.} 

For costs beyond sample complexity, the dominant classical computational cost is due to the mapping between the Ising theory and the LGT, which is linear in system size. Finally, the required circuit depth to implement the unitary rotations on the LGT side is quadratic in the system size due to the need for rotations generated by ribbon operators starting at the reference qubit $r$.

\section{Numerical Demonstrations}\label{s:example}
To enable the use and testing of our protocols, we provide Python code to perform the analysis for Global Dual Pairs and the Dual Product protocol~\cite{code}. To demonstrate the accuracy and sampling costs of the symmetry aware protocols (Global Dual Pairs and Dual Product) compared to the standard symmetry-ignorant Product protocol for estimating gauge-invariant observables, we numerically simulate the protocols applied to the ground state of the $\Ztwo$-LGT Hamiltonian in \cref{eq:H} for various system sizes and couplings $g$.
\begin{figure*}[t]
\centering\includegraphics[width=0.85\textwidth, trim={0 40 0 40}]{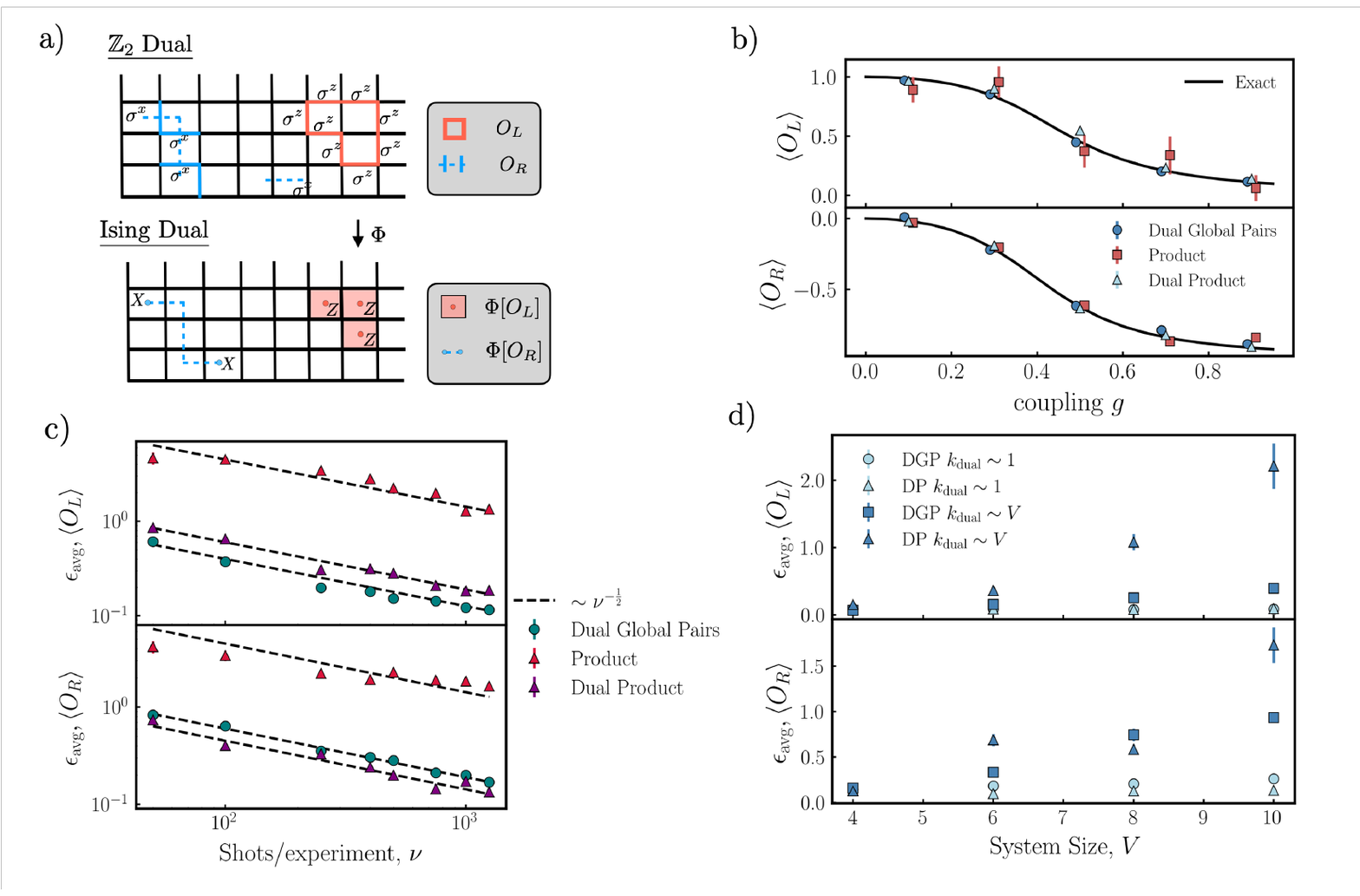}
\caption{\emph{Numerical Demonstration.} \textbf{(a)} Examples of a ribbon (blue) and loop (red) operator on both the $\mathbb{Z}_2$ LGT and Ising side of the duality. \textbf{(b)} Estimates of ribbon (of maximal length) and loop operator (two plaquettes) expectation values on a minimal lattice of size $N_x=3, N_y=2$ with PBC via Global Dual Pairs, Dual Product, and Product protocols, compared to the exact solution for the ground state of the $\Ztwo$ LGT as a function of the coupling $g$. Error bars represent the standard deviation of the median of means estimator over 50 experimental repetitions of $\nu=1000$ shots each.  \textbf{(c)} Average relative error $\epsilon_\mathrm{avg}$ (see \cref{eq:avg-rel-error}) versus number of samples $\nu$ for the same loop and ribbon operator as in panel (b) for the Product, Dual Product and Global Dual Pairs protocols for the ground state corresponding to coupling $g=0.5$. \textbf{(d)} Average relative error $\epsilon_\mathrm{avg}$ versus volume for loop and ribbon operators of fixed weight and operators with extensive weight for the Global Dual Pairs and Dual Product protocols. In particular, for the ribbon operators the extensive weights are 2,4,6, and 8 and for the loop operators they are 2,4,4, and 6, for systems sizes $V=4,6,7,10$. The average is taken over 50 experimental repetitions of $\nu=1000$ shots each using the ground state of the $\Ztwo$ LGT with coupling $g=0.5$. The sample cost of the Global Dual Pairs protocol depends both on the operator weight in the dual Ising theory and the size of the lattice, whereas the Dual Product protocol depends only on operator weight. 
}\label{fig:numerics-fig}
\end{figure*}
We estimate both ribbon operators $O_R$ with $\Phi[O_R]=X_iX_j$ and loop operators $O_L=\prod_{l\in L} \sigma^z_l$, where $L$ is a closed loop, as pictured in \cref{fig:numerics-fig}a. Our protocols accurately recover the expectation values of these observables, and perform a system size scaling analysis. 
As predicted by the analysis in the previous section, for observables with extensive non-trivial support on the LGT side of the duality, we observe exponential (in system size) sample complexity improvements of the Global Dual Pairs  versus the Product protocol. 

\cref{fig:numerics-fig}b shows a minimal example comparing Global Dual Pairs, Dual Product, and Product estimation for both a ribbon operator and loop operator as compared to the true values as a function of the coupling strength $g$ for a fixed system size $(N_x,N_y)=(3,2)$. We consider (i) a ribbon operator of the form $O_R=\sigma^x_i\sigma^x_j$ and (ii) a single plaquette loop operator of the form $O_L = \prod_{j\in\square} \sigma_j^z$.
The estimates for these example observables are computed using a fixed number of experimental shots $\nu=1000$ for all protocols. The plots show the mean and standard deviation of these estimates over 50 runs of the experiment. 
In each experiment, we employ a median-of-means estimator with block sizes of approximately $\sqrt{\nu}$, which is the standard choice in this setting as it suppresses the effect of outliers in the shadow table data~\cite{huang2020predicting}.\footnote{While we used median of means estimation for easy comparison with much of the classical shadows literature, we note that, consistent with Ref.~\cite{struchalin2021experimental}, empirically, we saw no real difference between median of means estimation compared to the simple sample mean.}
The true expectation values of the observables are also shown, and we see that all protocols provide accurate estimates with comparable standard deviation of the mean.
As expected, the error bars for a fixed number of shots for the Global Dual Pairs and Dual Product protocols compared to the Product protocol depends on the relative weight of the observables on the LGT side versus the Ising side of the duality. For instance, for the particular ribbon operator considered, on the Ising side of the duality the operator weight is $k_\mathrm{dual}=2$, whereas on the LGT side of the duality the operator weight is $k=3$, and, consequently, the error bars are slightly worse for the Product protocol than the symmetry-aware protocols. For the loop operator considered, the difference in error between the two protocols is more pronounced, as $k_\mathrm{dual}=2$ and $k=6$, and the error bars for the Product protocol are significantly larger for the same number of shots.

The expected behavior of the relative error for fixed samples is  shown in \cref{fig:numerics-fig}c, where we compare the Global Dual Pairs, Dual Product and Product protocols, showing the scaling of the average relative error over $M$ experimental repetitions
\begin{equation}    \label{eq:avg-rel-error}
\epsilon_\mathrm{avg}:=\frac{1}{M}\sum_{m=1}^M\frac{\left|\langle O\rangle_{\text{shadow},m}-\langle  O\rangle_{\text{exact}}\right|}{|\langle O\rangle_{\text{exact}}|},
\end{equation}  
as a function of the number of samples $\nu$ (per experiment) for fixed $V$ and fixed coupling $g=0.5$ for the same representative loop and ribbon operator as in \cref{fig:numerics-fig}b. Although all protocols exhibit the expected $1/\sqrt{\nu}$ shot noise scaling with sample number $N$, the Global Dual Pairs and Dual Product protocols achieve significantly lower errors than the Product protocol for any fixed $\nu$, in agreement with our analytical predictions.

As previously noted and further analyzed below, the relative accuracy of the  protocols depends on the operator weight of the observables being considered. To  demonstrate how the errors of the Global Dual Pairs and Dual Product protocols depend on the operator weight of the observables being estimated in the Ising picture, we compute the relative error averaged over $M=50$ experimental repetitions, $\epsilon_\mathrm{avg}$, as a function of system size $V$ for two different representative loop and ribbon operators,  shown in ~\cref{fig:numerics-fig}d. In particular, we show results for observables where the observable weight (on the Ising side) is constant and extensive in $V$. We use a fixed number of samples per experiment ($\nu=1000$) and fixed coupling $g=0.5$. 
The system sizes we can simulate numerically are too small to reliably verify the asymptotic scaling given by the error bounds of \cref{eq:var_gp}, but the qualitative behavior matches the expectations from our analytic results, namely, the cost of the Global Dual Pairs and Dual Product protocols increase with the Pauli weight of the observables on the Ising side of the duality. In the numerical examples shown, the size of systems for which we show data is limited by the classical emulation of the $\Ztwo$ LGT input state, not by the complexity of our protocols.

\begin{table*}[htb]
\centering
%\small
\renewcommand{\arraystretch}{1.}
\raggedright{\textbf{Arbitrary Gauge-Invariant Observables}}
\begin{tabular}{|m{4.45cm} >{\centering\arraybackslash}m{2.6cm} >{\centering\arraybackslash}m{4.6cm} >{\centering\arraybackslash}m{3.75cm}|}
\hline
\textbf{Protocol} & \textbf{Circuit Depth} & \textbf{Classical Cost per Sample} & \textbf{Number of Samples} \\
\hline

Global Dual Pairs (this work)     & $\mathcal{O}(V^2)$                 & $\mathcal{O}\left(\mathrm{poly}(V)\right)$ & $\mathcal{O}\left(\frac{1}{\epsilon^2} \mathrm{poly}(V)\right)$ \\
Dual Product (this work)     & $\mathcal{O}(V^2)$ & $\mathcal{O}(V)$& $\mathcal{O}\left(\frac{1}{\epsilon^2}\right)$  \\
Product~\cite{huang2020predicting}
& $1$                                & $\Theta(k) \subset \mathcal{O}(V)$              & $\mathcal{O}\left(\frac{1}{\epsilon^2} 4^k\right) \subset \mathcal{O}\left(\frac{1}{\epsilon^2} 4^V\right)$ \\
\hline
\end{tabular}
\vspace{0.5em}

\raggedright{\textbf{Geometrically Local, Gauge-Invariant Observables}}
\begin{tabular}{|m{4.45cm} >{\centering\arraybackslash}m{2.6cm} >{\centering\arraybackslash}m{4.6cm} >{\centering\arraybackslash}m{3.75cm}|}
\hline 
\textbf{Protocol} & \textbf{Circuit Depth} & \textbf{Classical Cost per Sample} & \textbf{Number of Samples} \\
\hline
Local Dual Pairs   (this work)    & $\mathcal{O}(L^4)$                 & $\mathcal{O}\left(\mathrm{poly}(V)\right)$ & $\mathcal{O}\left(\frac{1}{\epsilon^2} \mathrm{poly}(L)\right)$ \\
Dual Product (this work)      & $\mathcal{O}(V^2)$ & $\mathcal{O}(V)$& $\mathcal{O}\left(\frac{1}{\epsilon^2}\right)$  \\
Product~\cite{huang2020predicting}     & $1$                                & $\Theta(k) \subset \mathcal{O}(L^2)$            & $\mathcal{O}\left(\frac{1}{\epsilon^2} 4^k\right) \subset \mathcal{O}\left(\frac{1}{\epsilon^2} 4^L\right)$ \\
\hline
\end{tabular}
\caption{\emph{Summary of costs for estimating the expectation value of a gauge-invariant observable $O$ with $\norm{O}_\infty=1$ to an additive error $\epsilon$.} We assume that $O$ is $k$-local on the LGT side of the duality or $k_\mathrm{dual}$-local on the Ising side and, also, assume that $k_\mathrm{dual}\in\Theta(1)$. Recall that $V=N_xN_y$ is the size of the (dual) lattice and, for the Local Dual Pairs protocol, $L$ is the dimension is the size of the geometrically local patches considered. Circuit depths for implementing the random rotations in the Dual Pairs and Dual Product protocols are based on the standard decomposition of Pauli rotations into two qubit gates and assume no parallelization of the randomizing unitaries.}\label{tab:protocol-costs}
\end{table*}

\section{Protocol Comparisons}\label{s:comparisons}
The cost of shadows protocols are characterized in terms of their asymptotic performance for generic observables, typically depending on the Pauli weight of the target observable. For standard, symmetry-ignorant schemes, observables represented by constant-length Pauli strings are independent of system size, but for arbitrary Pauli strings, such as arise for generic gauge-invariant observables, the Pauli weight grows with system size. The sampling-cost reductions achieved by our protocols stem from the fact that, for a given target observable, the effective Pauli weight can differ non-trivially between the two sides of the duality. 
Here, we discuss the relative performance of the various protocols in more detail for specific classes of observables

\subsection{Comparing symmetry-aware protocols: Dual Pairs versus Dual Product Protocols}\label{s:pairs-vs-dual}
The Global Dual Pairs protocol and the Dual Product protocol accomplish the same goal, measuring \textit{any} gauge invariant observable.
The sample cost of the Dual Product protocol for gauge invariant observables with $k_\mathrm{dual}=\Theta(1)$ is significantly better, at least asymptotically, than the Global Dual Pairs protocol, i.e. constant as opposed to polynomial in system size.  This asymptotic gain is largely free: the classical costs per sample are both dominated by the mapping between the dual Ising theory and the LGT, yielding a polynomial in system size cost. The circuit depths for implementing the random unitaries are also both $\mathcal{O}(V^2)$. However, the constant factors are slightly better for the Global Dual Pairs protocol since, unlike the Dual Product protocol, the unitary rotations can be parallelized.\footnote{Parallelization is not possible for the Dual Product protocol because all rotations are generated by a ribbon operator that intersects with the reference plaquette $\square_r$ that defines the duality.}
Thus, since the Dual Product protocol suffers from only a constant factor worsening of circuit depth to move from polynomial to constant sample complexity, in the asymptotic regime it is simply the better protocol when compared to Global Dual Pairs. However, this  may not hold for small systems or for all observables.

When comparing the Dual Product and Local Dual Pairs protocols, and assuming geometrically local observables, the Local Dual Pairs protocol is superior. Even for geometrically local observables, the Dual Product scheme requires implementing rotations generated by ribbon operators extending to a reference plaquette, which prevents any reduction in the circuit depth of the randomizing unitaries. 
It is important to emphasize that these conclusions change in minor ways when we consider alternative boundary conditions; see for instance~\cref{a:global-FBC} for a discussion of fixed boundary conditions.

\begin{figure}[t]
\centering\includegraphics[width=0.67\columnwidth, trim = 50 180 50 180]{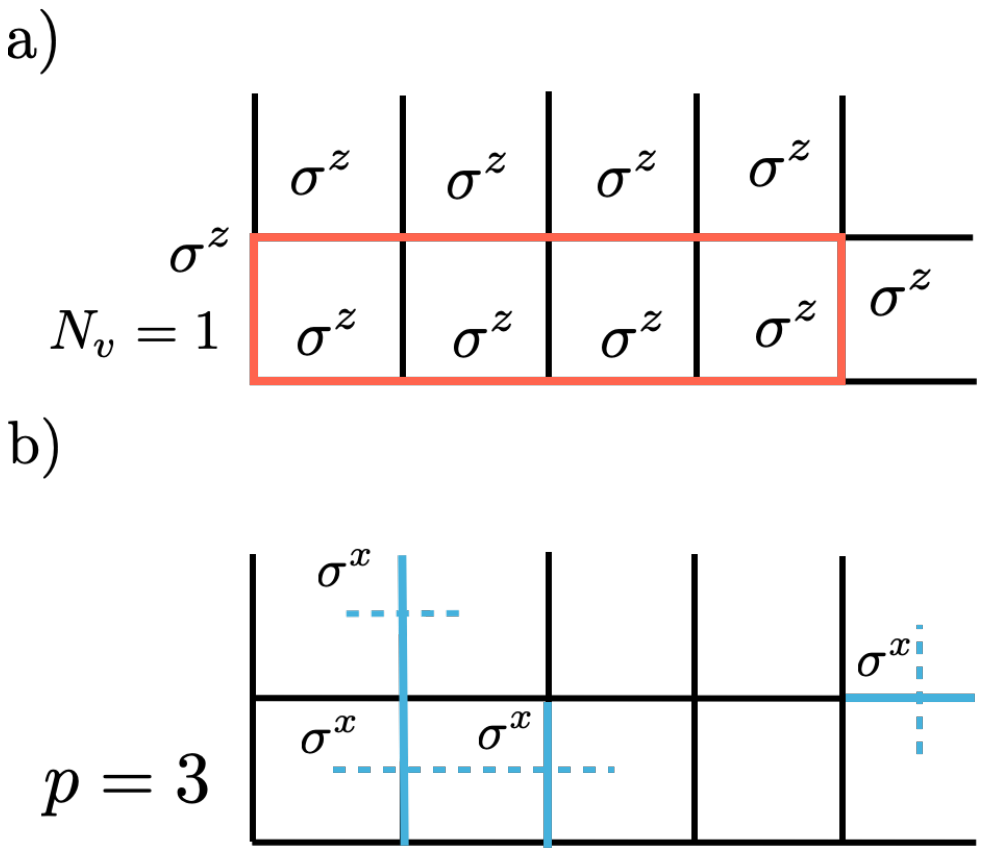}
\caption{\textbf{(a)} An illustration of an example loop operator of the sort considered in the discussion in \cref{s:compare-pauli}. \textbf{(b)}  An illustration of an example (set of) ribbon operators of the sort considered in the discussion in \cref{s:compare-pauli}. The way in which one picks paths to connect the end points is arbitrary, as all choices are equivalent up the action of Gauss law operators.}\label{fig:example-ops}
\end{figure}

\subsection{Comparing symmetry-aware to symmetry-ignorant protocols: Dual Pairs and Dual Product versus Product Protocols}\label{s:compare-pauli}
The standard approach for measuring $k$-local observables is the Product protocol, applied directly to the experimentally realized LGT system. 
Here, we compare this approach to the three symmetry-aware protocols developed in this work that make use of a virtual Ising system, 
without restrictions regarding the geometric locality of the observables of interest.

Compared to symmetry-aware protocols, the Product protocol offers a significantly lower circuit depth: it requires only a single layer of single-qubit unitaries, whereas, e.g., the Global Dual Pairs protocol involves $\Theta(V^2)$ high-weight Pauli rotations, each acting on $\mathcal{O}(V)$ qubits. However,  this simplicity comes at the cost of exponentially increased sample complexity for the worst case observables. To see the origin of the difference in sample complexity, recall that for the Product protocol, $\mathrm{Var}(o_j) \leq 4^k \norm{O}_\infty^2$,
 where $k$ is the locality of the operator in the LGT, and  $\norm{O}_\infty$ is evaluated on the full Hilbert space of the LGT~\cite{huang2020predicting} absent any gauge constraints. In contrast to the Product protocol, the Global Dual Pairs and Dual Product protocols have a sample cost that depends on the operator locality in the dual Ising theory, $k_\mathrm{dual}$.
 
In the worst case (for a symmetry-ignorant protocol), gauge invariant observables have $k\gg k_\mathrm{dual}$. For instance, for an observable of the form $X_i X_j$ (in the Ising picture), $k_\mathrm{dual}=2$, but the corresponding observable on the LGT side of the duality can have $k\sim V$ if the plaquettes $i,j$ are separated by a distance of the order of the system size. Consequently, for such observables, the Product protocol requires a number of samples exponential in the system size, whereas the Global Dual Pairs protocol requires only polynomial samples in the system size and the Dual Product protocol requires constant samples. Intuitively, these improvements are due to the exponential reduction in dimension from the full to the physical Hilbert space of the LGT. 
 
Beyond this extreme example, the difference in sample complexity between the Product protocol and the symmetry-aware protocols will depend on the particular observable(s) considered. To illustrate this, we examine how the locality of gauge-invariant observables differ on either side of the duality for several representative examples. In what follows, we let $w_i$ for $i\in\{X,Y,Z\}$ stand for the number of $\sigma^x$, $\sigma^y$, $\sigma^z$ operators, respectively, for some operator on the LGT side of the duality. Thus, $k=w_X+w_Y+w_Z$. Similarly, we let $\tilde{w}_i$ denote the same for $X$, $Y$, and $Z$ operators, respectively, on the Ising side of the duality, such that $k_\mathrm{dual}=\tilde{w}_X+\tilde{w}_Y+\tilde{w}_Z$. 

Our first example is the Wilson loop illustrated in \cref{fig:example-ops}a: a closed product of magnetic variables, $\prod_{\ell \in \mathcal{P}_{\circ}} \sigma^z_\ell$, which, in any Abelian LGT, can be equivalently expressed as the product of plaquette operators enclosed by the loop. We consider two limiting cases: (i) a rectangular Wilson loop of size $N_h \times N_v$, where $N_h$ and $N_v$ denote the horizontal and vertical extents of the loop, respectively; and (ii) a "line" of adjacent plaquettes of size $N_h' \times 1$, as illustrated in \cref{fig:example-ops}a. This may include the worst-case scenario where $N_h N_v \propto V$, the total system volume. The corresponding Pauli weights are as follows: for case (i), the LGT-side weight is $k = w_Z = 2(N_h + N_v)$, while the Ising-side weight is $k_\mathrm{dual} = \tilde{w}_z = N_h N_v$. Thus, for these observables, the Global Dual Pairs protocol yields a sampling advantage only when $N_h N_v < 2(N_h + N_v)$, which holds when either $N_h$ or $N_v$ is less than or equal to 2.  In contrast, for case (ii), the LGT-side weight is $k = w_Z = 2N_h' + 2$, while the Ising-side weight is $k_\mathrm{dual} = \tilde{w}_Z = N_h'$. Both examples show that symmetry-aware schemes are not universally optimal for all observables.

The large advantage of the Global Dual Pairs over the Product Protocol is driven mostly by ``electric'' observables, i.e., those involving electric operators $\sigma^x_\ell$ in the LGT ($X$ operators in the Ising dual). More generally,  consider observables that are products of $p$ pairs
of Pauli $X$ (and $Y$, which are products of electric and magnetic observables) operators on the Ising side, with $k_\mathrm{dual}=\tilde{w}_X+\tilde{w}_Y=2p$, see~\cref{fig:example-ops}b for illustration. For the case of $p$ pairs $X_i X_j$, the mapping to an operator in the LGT is not unique---all give the same expectation value in the gauge invariant sector. In practice, on the LGT side, one would optimize by finding a collection of configurations with minimal cumulative Manhattan distance, resulting in a weight $k=\sum_{\{ l_a\}_{a\le p} }l_a$, while the Ising side weight is simply the smaller $p$. Considering $Y$ operators corresponds to products of electric and magnetic operators in the LGT with an even greater Pauli weight, while the corresponding weight in the Ising dual stays the same. 

This shows that, unsurprisingly, the largest \textit{individual} gains are observed for electric gauge-invariant observables rather than magnetic ones, since the Gauss laws—responsible for reducing the physical Hilbert space—are defined in the electric basis~\footnote{One could therefore argue that no advantage exists if measurements are restricted to magnetic observables only. However, in that case, any random measurement protocol would be entirely unnecessary, as all such observables commute and could simply be measured in the same fixed basis.}.
For context,  the key advantage of classical shadows is the ability to estimate \emph{multiple} (non-commuting, i.e., with different eigenbasis) observables with the same samples. Thus, while sample complexity improvements may not be present for selected individual observables, provided we want to be able to estimate many arbitrary gauge invariant observables the upper bounds on sample complexity that we provide in \cref{tab:protocol-costs} accurately reflect the sample complexity for estimating $M$ arbitrary gauge-invariant observables, up to the $\log(M)/\epsilon^2$ factor in \cref{eq:nubound}.

The classical computational cost (per sample) of inverting the shadow channel is comparable between the Product protocol, and the symmetry-aware Global Dual Pairs and Dual Product protocols. For both protocols the per sample classical cost is dominated by the $\in\mathcal{O}(\mathrm{poly}(V))$ due to computing the mapping from the measurement outcomes to dual bit strings (i.e. the mapping $\vec{s}\rightarrow\vec{b}$). For Global Dual Pairs, one must also pre-compute $f$ and $\alpha$ for a given observable, as described in \cref{s:global-protocol}. An overall comparison of the (worst case) resource costs for the Product, Dual Product, and Dual Pairs protocols is given in \cref{tab:protocol-costs}.

\section{Discussion and Outlook}\label{s:discussion}
In this manuscript, we introduced three variants of randomized measurement schemes tailored for systems with local (gauge) symmetries. 
We analyzed their use for estimating gauge-invariant observables in $\mathbb{Z}_2$ LGT, leveraging a LGT–Ising duality which enables detailed  analysis and the derivation of formal guarantees. We systematically determined, for any given set of observables, a practical advantage over symmetry-agnostic alternatives, and provide explicit algorithms supported by numerical implementations~\cite{code}. 

As with most protocols that make use of prior information---in this case, local (gauge) symmetry---a tradeoff between a reduction in sampling and an increase in circuit complexity arises. For NISQ-era devices, circuit complexity is most often quantified by CNOT depth: while the standard Product protocol~\cite{huang2020predicting}, can be achieved in constant depth, the (Global) Dual Pairs and Dual Product protocols we propose require entangling (although still shallow) circuits. For instance, for the Global Dual Pairs protocols, the required unitary rotations for a given pairing requires a depth of $2N-2$, where $N$ is the number of plaquettes between any two pairs, implying a total CNOT depth of $\mathcal{O}(V^2)$, shown in Table~\ref{tab:protocol-costs}. Similar circuit depths are required for the Dual Product protocol due to the need for ribbons going to the reference plaquette $\square_r$ for all rotations. For geometrically local observables, this  can be reduced by the Local Dual Pairs protocol to $\mathcal{O}(L^4)$, while maintaining an exponential improvement (in $L$) in sample complexity over the standard Product protocol. In the near-to-intermediate term whether or not the polynomial increase in circuit depth associated with tailoring shadow protocols is worth the exponential gains in sample complexity will depend on the platform and the ability to implement circuits of sufficient depth without incurring large errors due to imperfect gates. 
However, in fault-tolerant quantum processors---assuming that the timescales for implementing logical gates are comparable to or better than those for measurement and readout---the potential exponential reduction in sample complexity constitutes an unalloyed advantage when simulating $\mathbb{Z}_2$ LGT directly.

Nonetheless, in practice, the same Ising-LGT duality that allows us to rigorously analyze the performance of the shadows protocols could be used to directly simulate the dual Ising model~\cite{mueller2025quantum} instead of the LGT, and avoid the large circuit depth overheads required. In this case, the sample complexity for our shadows schemes would be identical, but the circuit depths would be significantly smaller. Our work, thus, should not be primarily understood as an explicit simulation protocol for $\mathbb{Z}_2$ LGT, but rather the key components are (a) a rigorous analysis of a constructive shadows protocol for an explicit model, demonstrating the circuit depth-sample complexity trade-off intrinsic to symmetry-preserving shadows ~\cite{sauvage2024classical} that is expected for LGTs; and (b) a blueprint for analyzing more general LGTs, including non-Abelian LGTs relevant to high energy and nuclear physics such as SU(2) and SU(3) LGTs even if dualities may be hard to find or do not exists.

Indeed, while the central insight of our work relies on the Ising–LGT duality specific to $\mathbb{Z}_2$ (and by extension $\mathbb{Z}_N$, and in the appropriate limit, $U(1)$~\cite{horn1979hamiltonian}) LGTs, many elements of our protocols generalize naturally. All steps of the protocols considered can be cast fully in the language of the LGT, rather than the dual Ising model. In this sense, the $\mathbb{Z}_2$ LGT case serves as a useful starting point, as the presence of a duality allows for a rigorous
analysis of the channel and shadow protocol. Building on this foundation, one can move beyond this simple setting by using these insights as a blueprint for more general LGTs---particularly to \emph{non-Abelian} LGTs or models whose local Hilbert spaces do not resemble qubits (including LGTs based on compact groups)---where the corresponding gauge-invariant
randomizing operations follow directly. However, what channels these randomizing operations represent and if (and how efficient) they can be inverted is remains a major challenge. 
We presently are not aware of a simple analogous approach for non-Abelian LGTs for which dualities, so far, are challenging to construct~\cite{handler1978duality,drouffe1979lattice,alvarez1994non,buchstaber2003generalized,cobanera2013solution,ji2020categorical,aasen2020topological}. This raises the question of whether similar sampling-efficient protocols can be devised for arbitrary LGTs, and whether their existence implies an underlying Ising-like duality, or their absence indicates its nonexistence. An interesting development in this context is the recent demonstration of charge-singlet measurements in quantum simulations of $SU(N_c)$ LGT~\cite{chakraborty2025charge,than2024phase}. In the absence of a duality, one could study the corresponding channels numerically in small-scale instances, but extracting scaling from this is a challenge. Overcoming these hurdles is a primary goal for future work.

A variety of alternative approaches to symmetry-respecting shadows have been recently proposed. One promising direction~\cite{zhan2024learning} employs classical shadows to learn unknown symmetries of quantum systems, potentially offering a potential "bootstrap" method for constructing symmetry-respecting classical shadows. Ref.~\cite{sauvage2024classical} introduces a general framework to construct symmetry-respecting shadows under the assumption of oracle access to a unitary that block diagonalizes the system in the irrep basis of the associated symmetry. Another established approach uses symmetric approximate unitary $k$-designs for LGTs~\cite{bringewatt2024randomized}, which allow simple channel inversion once the symmetry-sector dimensions are known but require deep random circuits. Recent results on log-depth approximate unitary $k$-designs could extend to random circuits with symmetries~\cite{ji2018pseudorandom,ma2025construct,schuster2025random}. Applying such constructions to systems with gauge symmetries is likely to be highly favorable when estimating properties of the state beyond $k$-local observables~\cite{grevink2025will,cui2025random}. Finally, there exist other symmetry-ignorant shadow tomography schemes that occupy an intermediate regime between the Product and Dual Pairs protocols in terms of both sampling cost and circuit complexity. Notable examples include the locally entangled schemes of Ref.~\cite{ippoliti2024classical} and the shallow shadow schemes of Refs.~\cite{hu2023classical,hu2025demonstration}. The potentially exponential advantages observed in our setting stem from explicitly exploiting gauge symmetry and, thus, will extend to these settings.

Significant progress has been made in understanding  the impact of noise on standard schemes~\cite{chen2021robust,koh2022classical,rozon2024optimal,brieger2025stability}, and the development of derandomized variants~\cite{huang2021efficient}. These ideas are likely extendable to the symmetry-respecting circuits. However, a particularly interesting complication arises: noise can break the very symmetries that these protocols are designed to exploit. This symmetry breaking can lead to biased estimators and necessitates a fundamentally different error analysis. Relatedly, the impact of small gauge violations in the input state on the performance of our protocols remains an interesting open question. On the other hand, symmetry-aware measurement schemes can enable powerful error mitigation schemes, as they allow to effectively project the measured state in to a symmetric subspace \cite{hu2022logical}.

\begin{acknowledgments}
We thank Hong-Ye Hu, Jonathan Kunjummen, Martin Larocca, and Yigit Subasi for helpful discussions. J.B. thanks the Harvard Quantum Initiative for support. H.F. and N.M. (during early stages) acknowledge funding by the DOE, Office of Science, Office of Nuclear Physics, IQuS (\url{https://iqus.uw.edu}), via the program on Quantum Horizons: QIS Research and Innovation for Nuclear Science under Award DE-SC0020970. J.B. notes that the views expressed in this work are those of the author and do not reflect the official policy or position of the U.S. Naval Academy, Department of the Navy, the Department of Defense, or the U.S. Government.
\end{acknowledgments}

\bibliography{bibi.bib}

\clearpage
\onecolumngrid
\appendix

\section{Details of Sample Complexity Bound}\label{a:details-sample-complexity}
Here, we provide details for the derivation of the bound on sample complexity for the Global Dual Pairs protocol for $\Ztwo$ LGT with PBC. In particular, we provide a derivation of \cref{eq:var_gp}. From Lemma S1 of \cite{huang2020predicting}, for a traceless operator $O$, the variance of the corresponding shadow expectation value $o_j$, where  $\mathbb{E}_{U,\mathbf{b}}(o_j)=\text{Tr}(\rho O)$,  
\begin{align}
\mathrm{Var}(o_j)=\mathrm{Var}\left(\mathrm{Tr}\left[\widehat{\Phi[\rho] }O\right]\right)\leq \norm{O}_\mathrm{shadow}^2,
\end{align}
where $\rho$ is the unknown quantum state and, thus, $\widehat{\Phi[\rho]}$ is the Ising-side classical shadow of this state. The so-called shadow norm is defined as
\begin{align}
\norm{O}_\mathrm{shadow}^2 = \max_{\sigma}\, \mathbb{E}_{U}\sum_{\vec{b}\in\{0,1\}^n}\bra{\vec{b}}U\sigma U^\dagger\ket{\vec{b}}\bra{\vec{b}}U\mathcal{M}^{-1}[O] U^\dagger\ket{\vec{b}}^2,
\end{align}
where $U$ are the random unitaries applied before measurement, $\vec{b}$ are the measurement outcomes (i.e. bit strings), $\sigma$ is some quantum state, and $\mathcal{M}$ is the shadow channel (see \cref{eq:channel}). 
To apply these expressions, we must consider the operator $O$ on the Ising side of the duality. 

Without loss of generality, assume that $O$ is a single Pauli string. Then, from \cref{eq:channel-factorization}, $\mathcal{M}$ acts proportionally to identity and is trivial to invert yielding:
\begin{align}
\mathcal{M}^{-1}(O) = \frac{3^{w_{xy}/2}}{f\alpha} O,
\end{align}
where we recall that  $w_{xy}$ is the total number of $X$ and $Y$ operators in the (Ising-side) operator string $O$, $f$ is the fraction of pairings $\pi\in P_V$ for which $\mathcal{M}_\pi(O)\neq 0$ (\cref{eq:f}), and $\alpha$ is an amplitude associated with permutations of $Z$ and $I$ operators within the operator $O$ (\cref{eq:aLocal Pairsha}).
Thus, the shadow norm is  bounded as
\begin{align}\label{eq:shadow-app}
\norm{O}_\mathrm{shadow}^2&= \max_\sigma \mathbb{E}_{U,\vec{b}}\bra{\vec{b}}U\mathcal{M}^{-1}[O] U^\dagger\ket{\vec{b}}^2 \leq \frac{3^{w_{xy}/2}}{f\alpha} \norm{O}_\infty^2,
\end{align}
where we use that $\bra{\vec{b}}UO U^\dagger\ket{\vec{b}}^2\leq \lambda_\mathrm{max}(O)^2$, where $\lambda_\mathrm{max}(O)$ is the largest eigenvalue of $O$. Substituting this bound on the shadow norm into \cref{eq:shadow-app} yields \cref{eq:var_gp}.

\section{Protocols with Fixed Boundary Conditions}\label{a:global-FBC}

In this Appendix, we explicitly describe the duality with fixed boundary conditions (FBC) (\cref{a:FBC-duality}), i.e. where the electric flux entering the system is fixed, and explain in detail the Global Dual Pairs protocol in this formulation, including a numerical example (\cref{a:global-pairs-FBC}). The Local Dual Pairs and Dual Product protocols can be  generalized along similar lines.

\begin{figure}[htb]
\centering\includegraphics[width=0.95\textwidth,trim = 0 50 0 80]{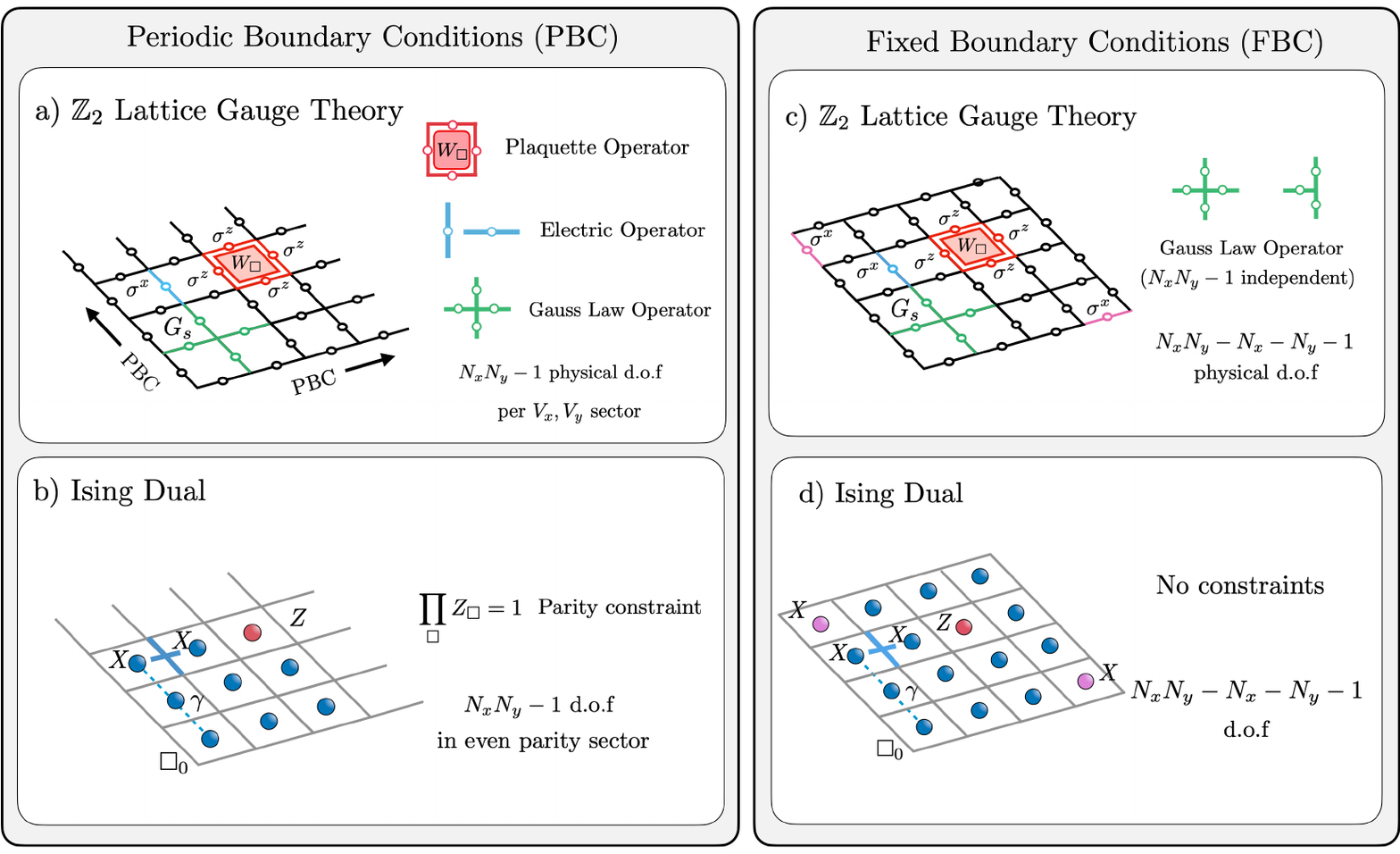}
\caption{\emph{Summary of LGT-Ising duality for PBC and FBC.} (a) Hilbert space structure and operators for $\mathbb{Z}_2$ LGT with PBC and (b) for the corresponding Ising dual which is subject to a parity constraint. Spheres mark one local spin-1/2 d.o.f. (c) For FBC, the LGT includes bulk and boundary Gauss laws, as well as plaquette and electric operators. (d) The dual Ising theory for FBC is no longer parity constraint but electric operators dualize differently in bulk and on the boundary. }\label{fig:FBC}
\end{figure}

\subsection{Duality with FBC}\label{a:FBC-duality}
We now describe how the duality is modified in the presence of fixed boundary conditions (FBC). The case of periodic boundary conditions (PBC) is discussed in the main text and, for reference, is also summarized in \Fig{fig:FBC}a and b. The LGT–Ising duality under FBC is illustrated in \Fig{fig:FBC}c (LGT side) and d (Ising side).

A key difference from the PBC case is the absence of global superselection operators ($V_x$, $V_y$). With FBC, there are $N_x(N_y - 1) + N_y(N_x - 1)$ qubit degrees of freedom residing on the links of the lattice, and $N_xN_y - 1$ independent Gauss-law constraints. The subtraction by one arises from the identity $\prod_s G_s = 1$, which includes both bulk and boundary sites $s$ of the lattice. As a result, the number of remaining physical degrees of freedom is $N_xN_y - N_x - N_y + 1$.
This matches the number of degrees of freedom in the dual Ising system, which consists of $(N_x - 1)(N_y - 1)$ dual spins located on the plaquettes. The explicit operator mapping is detailed in \Fig{fig:FBC}c and d. For clarity, we label operators as belonging to the bulk or the boundary, depending on their location. The electric variables in the bulk, $\sigma^x_l$, and all magnetic terms $W_\square$ map to the dual theory just as in \cref{eq:duality-map}. However, electric operators $\sigma^x_l$ on the boundary map to a single $X_\square$ operator. That is,
\begin{equation}\label{eq:dual-map-bndry}
\Phi(\sigma^x_l) = X_\square, \quad (\text{for } l \text{ on boundary}),
\end{equation}
where $\square$ is the single plaquette adajcent to the boundary link $l$; see \cref{fig:FBC}b for an illustration.
\Cref{eq:dual-map-bndry} implies that for $\sigma^x_l$ operators located at the lattice corners, the two electric fields meeting at a corner both map to the same dual operator $X_\square$. This reflects that the corner $\sigma^x_l$ operators are equivalent up to the action of the local Gauss law operator.
The dual Hamiltonian for $\mathbb{Z}_2$ LGT in 2+1d with FBC can be written as,
\begin{align}\label{eq:dualHFBC}
H_\dual^\mathrm{FBC} = -\sum_\square Z_\square -g\sum_{l \in\mathrm{boundary}} X_{\eta_l} \nonumber -g\sum_{\square\in\mathrm{blk}} (X_\square X_{\square-\hat{x}}+X_\square X_{\square-\hat{y}}),
\end{align}
where $\eta_l$ runs over plaquettes adjacent to $l$, $\square\cap l\neq\emptyset$. Unlike the PBC case, there is no parity symmetry constraint in the dual, matching with the fact that on the LGT side it is no longer true that $\prod_\square W_\square=1$. 

\subsection{Adapting the Global Dual Pairs Protocol to FBC}\label{a:global-pairs-FBC}
The Global Dual Pairs protocol with FBC is essentially identical to the PBC case discussed in the main text, except that the parity symmetry constraint is no longer imposed. Accordingly, for each pairing $\pi$, and for pairs $[ij]\in\pi$, we apply a randomizing unitary $U_{[ij]}$ drawn from a full two-qubit (approximate) unitary 2-design, i.e., without enforcing parity symmetry. For these pairs~\cite{mele2024introduction},
\begin{equation}\label{eq:two-design-action-FBC}
\mathcal{M}_{[ij]}\big[\Phi[\rho]\big]=\frac{1}{5}\Big(\Phi[\rho]+\mathrm{Tr}\big[\Phi[\rho]\big]\Big).
\end{equation}
Thus, the action on an operator string $S_iS_j\in\{I,X,Y,Z\}^{\otimes 2}$ is
\begin{align}\label{eq:channel-action-FBC}
&\mathcal{M}_{[ij]}[S_iS_j]=\begin{cases}
S_iS_j, & S_iS_j=II\\\
\frac{1}{5} S_iS_j,  & \text{otherwise}.
\end{cases}
\end{align}
Consequently, $\mathcal{M}$ is diagonal and easily invertible. That is,
\begin{equation}
\mathcal{M}[S]=\tilde{\alpha} S,
\end{equation}
where $\tilde{\alpha}$ is the amplitude associated with $S$, defined as
\begin{align}
\tilde\alpha = \frac{1}{|P_V|}\sum_{m=0}^{\lfloor k_\mathrm{dual}/2\rfloor} \frac{1}{5^{k_\mathrm{dual}-m}}\underbrace{\binom{k_\mathrm{dual}}{2m} |P_{2m}|}_{\text{double pairings}}\underbrace{\binom{V-k_\mathrm{dual}}{k_\mathrm{dual}-2m}(k_\mathrm{dual}-2m)!}_{\text{single pairings}}
\end{align}
where $k_\mathrm{dual}$ is the number of non-identity operators in $S$ (i.e. the Pauli weight of $S$), and where the sum is over the number of pairings $m$ that pair non-trivial operators with one another (``double pairings'').  Unlike the PBC case, all pairings contribute and, thus, there is no FBC equivalent to the fraction $f$ in \cref{eq:channel-factorization,eq:f}. 
Pairings that pair a single non-trivial operator with an identity operator are denoted a ``single pairing.'' Assuming that $k_\mathrm{dual}=\Theta(1)$, and, using Stirling's approximation we find that $\tilde{\alpha} \sim 1/\mathrm{poly}(V)$. Thus, following a computation nearly identical to that in \cref{a:details-sample-complexity}, the sample cost remains polynomial in the system size, as in the PBC case.

\begin{figure}
    \centering
    \includegraphics[scale=.5, trim = 0 150 0 120]{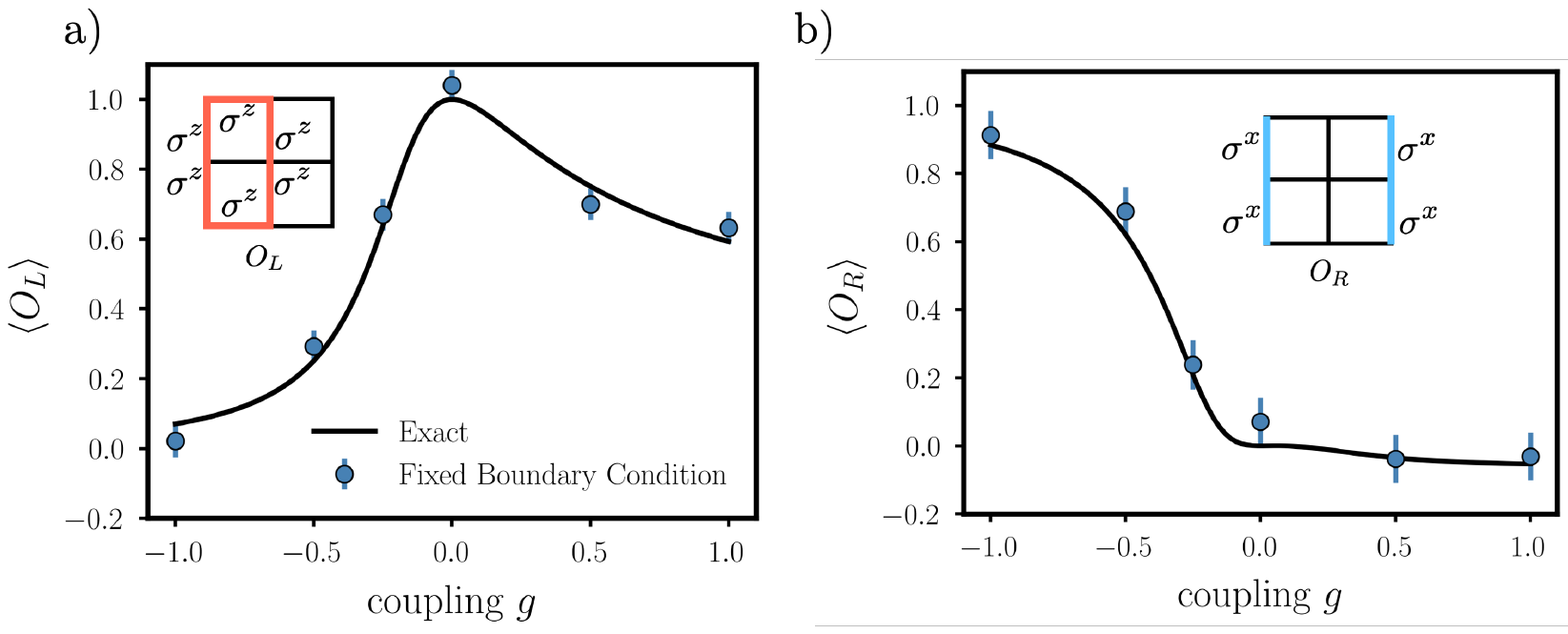}
    \caption{\textit{Implementation of Global Dual Pairs with Fixed Boundary Conditions} Example of \textbf{a)} loop and \textbf{b)} ribbon operators for $N_x=3$ and $N_y=3$ with FBC via the modified Global Dual Pairs protocol, compared to the exact solution of the $\mathbb{Z}_2$ LGT ground state. Error bars represent standard deviation of the mean over a single experiment of $\nu=5000$ shots.}
    \label{fig:fbc_plots}
\end{figure}

\Cref{fig:fbc_plots} presents a numerical demonstration of the Global Dual Pairs protocol with FBC, implemented on a $(N_x, N_y) = (3,3)$ lattice. We estimate a loop operator $O_L$ in (a), and a 
ribbon operator $O_R$ in panel (b). The FBC adaptation accurately reproduces the expected values across a broad range of coupling strengths. In addition, the convergence to the exact result is quicker for the loop observable $O_L$ compared to the ribbon operator $O_R$, which while having higher weight on the $\mathbb{Z}_2$ side of the duality, has lower weight compared to $O_R$ on the Ising side.

\end{document}